\documentclass[aps,pra,superscriptaddress,twocolumn]{revtex4}
\usepackage{epsfig}
\usepackage[usenames]{color}
\usepackage{amssymb,amsmath,bbm,amsfonts,amsthm,subfigure}

\newcommand{\ket}[1]{\left\vert#1\right\rangle}

\newcommand{\bra}[1]{\left\langle#1\right\vert}

\newcommand{\valmed}[1]{\left\langle#1\right\rangle}

\def\Tr{\mbox{Tr}}

\begin{document}

\title {Work statistics, irreversible heat and correlations build-up in joining two spin chains}

\author{T.~J.~G. Apollaro}
\affiliation{Dipartimento di Fisica \& INFN--Gruppo collegato di Cosenza, Universit\`a della Calabria,
       Via P.Bucci, 87036 Arcavacata di Rende (CS), Italy}
\affiliation{Centre for Theoretical Atomic, Molecular, and Optical Physics,
   School of Mathematics and Physics, Queen's University Belfast,
   BT7\,1NN, United Kingdom}
   
   \author{Gianluca Francica}
\affiliation{Dipartimento di Fisica \& INFN--Gruppo collegato di Cosenza, Universit\`a della Calabria,
       Via P.Bucci, 87036 Arcavacata di Rende (CS), Italy}
   
   \author{Mauro Paternostro}
\affiliation{Centre for Theoretical Atomic, Molecular, and Optical Physics,
  School of Mathematics and Physics, Queen's University Belfast,
 BT7\,1NN, United Kingdom}

   \author{Michele Campisi}
\affiliation{NEST, Scuola Normale Superiore \& Istituto di Nanoscienze-CNR, I-56126 Pisa, Italy}

\begin{abstract}
We investigate the influences of quantum many-body effects, such as criticality and the existence of factorisation fields, in the thermodynamic cost of establishing a bonding link between two independent quantum spin chains. We provide a physical interpretation of the behavior of irreversible work spent in such process by linking the phenomenology of such quantities to the properties of the spectrum of the system.
\end{abstract}
\maketitle

%%%%%%%%%%%%%%%%%%%%%%%%%%%%%%%%%%%%%%%%%%%%%%%%%%%%
%%%%%%%%%%%%%%%%%%%%%%%%%%%%%%%%%%%%%%%%%%%%%%%%%%%%
\section{Introduction}
%%%%%%%%%%%%%%%%%%%%%%%%%%%%%%%%%%%%%%%%%%%%%%%%%%%%
The statistics of work in quantum systems \cite{Esposito09RMP81,Campisi11RMP83,Tasaki00arxiv,Kurchan00arxiv,Talkner07PRE75,Campisi09PRL102,Talkner09JSM09} subjected to a time dependent processes has recently attracted a remarkable body of work~
\cite{Pekola12arxiv,Dorner13PRL110,Mazzola13PRL110,Campisi13NJP15,Mazzola14arxiv} 
%\cite{Pekola12arxiv,Kafri12PRA86,Abah12PRL109,Ngo12PRE86,Albash13PRE88,Dorner13PRL110,Mazzola13PRL110,Mazzola14arxiv,Goold14arxiv,Goold14arxiv2} 
aimed at predicting the behaviour of thermodynamically relevant quantities (work, free-energy variations, and entropy) in finite-time dynamics. 
Experimental efforts in both the classical and quantum mechanical scenarios~
\cite{Liphardt02SCIENCE296, Collin05NAT437, Douarche05EPL70, Toyabe10NP6,Saira12PRL109, Batalhao13arxiv} have started burgeoning, showing the concrete possibility to test the validity of fluctuation theorems~\cite{Crooks99PRE60,Jarzynski97PRL78,Esposito09RMP81,Campisi11RMP83}. Although, with a few exceptions~\cite{Silva08PRL101,Dorner12PRL109,Smacchia13PRE88,Sindona13PRL111,Mascarenhas13arxiv,Sindona14NJP16,Carlisle14arxiv,Fusco14arxiv}, the efforts so far have been concentrated on the single and/or few-body case, there is natural interest in extending such investigations to the quantum many-body domain, whose rich physics would offer new chances to investigate non-equilibrium quantum thermodynamics. 

In this paper, we contribute to this ongoing effort by providing an extensive study of the out-of-equilibrium thermodynamics of a process consisting in the sudden establishment of a bond between two otherwise mutually disconnected finite-size spin chains and linking the corresponding phenomenology to the many-body properties of the system. In particular, we concentrate on features of criticality and factorizability of the spin models at hand, highlighting the connections with the amount of irreversible work produced as a result of the sudden bonding of the two chains, which indeed exhibits features that are reminiscent of the properties of the spectrum of the model being studied.

The remainder of this manuscript is organized as follows: In Sec.~\ref{tools} we introduce the instruments that will be used in our analysis of the system that is described in Sec.~\ref{S.Model}. We will highlight that the irreversible work can be also understood as the irreversible heat that the non-equilibrium system will cede if at the end of the driving the system undergoes a re-thermalization step. The statistics of work produced in the sudden bonding process is analysed in Sec.~\ref{work} and further illustrated, with special emphasis given to the irreversible version of work due to the non-adiabatic nature of the process at hand, in Sec.~\ref{S.quenchvsh}. The link with the features of the spectrum of the model, from criticality to factorisation, is then provided in Secs.~\ref{sS.equalNhfact}, \ref{sS.difN}, and \ref{sS.equalN}.

\section{nonequilibrium lag, irreversible work and irreversible heat}
\label{tools}

%%%%%%%%%%%%%%%%%%%%%%%%% FIGURE %%%%%%%%%%%
\begin{figure}[]
		\begin{center}
		\includegraphics[width=\linewidth]{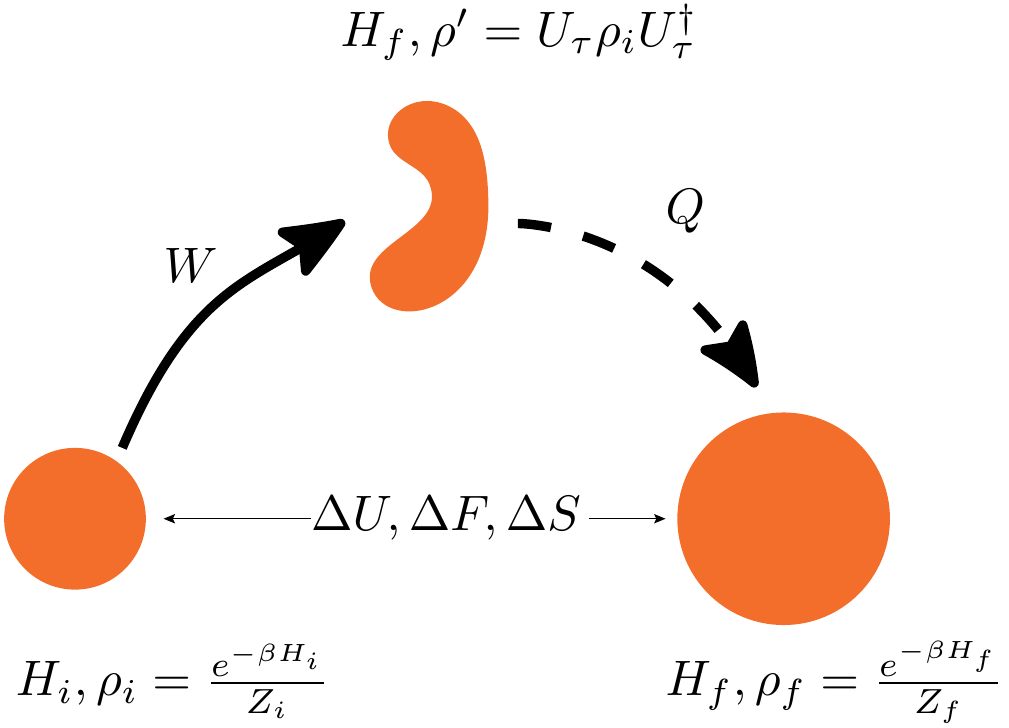}
		\caption{(Color Online) The initial thermal equilibrium state $\rho_i$ (small circle) evolves unitarily (solid arrow)
		into the non-equilibrium state $\rho'$ (bean-shaped) in response to an external driving changing the Hamiltonian from $H_i$ to $H_f$. During this step the average work $W$ is performed and no heat is exchanged (the evolution occurs in thermal isolation). The non-equilibrium state $\rho'$ evolves into the final equilibrium $\rho_f$ (large circle) during the non-unitary thermalization evolution. During this step the heat $Q$ is exchanged with the bath and no work is performed. }
		\label{fig:Fig1}
		\end{center}
\end{figure}
%%%%%%%%%%%%%%%%%%%%%%%%%%%%%%%%%%%%%%%%%%
We consider the following general set-up illustrated in  Fig. \ref{fig:Fig1}.
At time $t=0$ a system with initial Hamiltonian $H_i$ stays in thermal equilibrium
with a bath of temperature $T=1/k_B \beta$ 
\begin{equation}
H(0) = H_i \, ,\quad \rho(0)= \frac{e^{-\beta H_i}}{Z_i} := \rho_i
\end{equation}
where $k_B$ denotes Boltzmann's constant.
At time $t=0^+$ the system is detached from the bath and evolves unitarily until time $t=\tau$ under the action of a time dependent forcing specified by the time-dependent system Hamiltonian $H(t)$, taking the Hamiltonian from $H_i$ to $H_f$:
\begin{equation}
H(\tau) = H_f \, ,\quad \rho(\tau)= U_\tau^\dagger \rho_i U_\tau := \rho'
\end{equation}
Here $U_t$ is the evolution generated by $H(t)$, namely $i\hbar \dot U_t = H(t) U_t$.
At time $t=\tau$ connection with the bath is re-established and after a sufficient long time $\mathcal{T}$, 
the system finally reaches the state of thermal equilibrium $\rho_f$
\begin{equation}
H(\tau+\mathcal{T}) = H_f \, ,\quad \rho(\tau+\mathcal{T})= \frac{e^{-\beta H_f}}{Z_f} := \rho_f
\end{equation}
We will consider this unitary+thermalization process as the basic building block of a generic non-equilibrium thermodynamic
process, very much in the spirit of Refs.~\cite{Crooks08JSM08, Anders13NJP15}.

As a consequence of the work fluctuation relation \cite{Campisi11RMP83}
we have that the average work $\langle w \rangle= \Tr (\rho'H_f -\rho_i H_i)$ done on the system during the time $0<t<\tau$ is larger than 
the free energy difference $\Delta F = -\beta^{-1}\ln Z_f/Z_i$ between the equilibrium states $\rho_f$ and $\rho_i$
\begin{equation}
 \langle w \rangle -\Delta F \geq 0.
\label{eq:W-DF}
\end{equation}
This difference is often referred to as the irreversible work $W_\text{irr}$ and quantifies the lag between
the nonequilibrium state $\rho'$ and the corresponding equilibrium $\rho_f$ \cite{Vaikuntanathan09EPL87}
\begin{equation}
W_\text{irr} =  \langle w \rangle -\Delta F = \beta^{-1} D[\rho'||\rho_f].
\label{eq:KL}
\end{equation}
where $D[\rho||\sigma]= \Tr{\rho \ln \rho}- \Tr {\rho \ln \sigma} $
denotes the Kullback-Leibler divergence, which quantifies the degree of indistinguishability
between $\rho$ and $\sigma$. 

The non-equilibrium lag can also be interpreted in a different way if one focusses
on the thermalization step:
Consider the difference in thermodynamic internal energy $\Delta U = \Tr \rho_f H_f - \Tr \rho_i H_i$
of the two equilibrium states $\rho_f$ and $\rho_i$. On one hand, using standard
canonical ensemble manipulations one can write it as 
\begin{equation}
\Delta U = \Delta F + T \Delta S
\end{equation}
where $\Delta S = -\Tr \rho_f \ln \rho_f + \Tr \rho_i \ln \rho_i$.
On the other hand, by adding
and subtracting $ \Tr \rho' H_f $ one has
\begin{align}
\Delta U &=  \Tr (\rho_f H_f - \rho' H_f )+  \Tr (\rho'H_f -\rho_i H_i)\\
&= \langle Q \rangle + \langle w \rangle \label{eq:QW}
\end{align}
where, under the usual assumption of weak system-bath coupling, the quantity  
$ \Tr (\rho_f H_f - \rho' H_f )$, i.e. the energy gained by the system
during the thermalization step, $\tau<t<\tau+\mathcal{T}$, is interpreted as heat.
Combining these expressions together we arrive at
\begin{align}
\langle w \rangle -\Delta F  = T \Delta S - \langle Q \rangle.
\end{align}
Eq. (\ref{eq:W-DF}) then implies $\langle Q \rangle \leq T\Delta S$, which is Clausius formulation of the second law, saying that the heat ceded
to the bath during the equilibration step is smaller than the 
heat $T\Delta S$ that would have been given to the bath if the system
was brought from $\rho_i$ to $\rho_f$ through a quasi-static isothermal transformation.
The latter can be approximately be reached by means of many
small quenches followed each by a thermalization step of duration
$\mathcal{T}$ or larger. In the limit of infinitely many infinitesimal such quench+thermalization steps,
the equality $ \langle Q \rangle = T\Delta S $ would be reached.

In the following we shall call the difference $\langle Q \rangle - T\Delta S  $ the irreversible 
heat $Q_\text{irr}\leq0$. It is given by
\begin{align}
 Q_\text{irr}=  -W_\text{irr} = -\beta^{-1} D[\rho'||\rho_f].
 \end{align}
For an arbitrary $\rho_i$, the above formulation of the second law implies the formula
$\beta\Tr (\rho_f-\rho_i)H_f \leq -\Tr \rho_f \ln \rho_f + \Tr \rho_i \ln \rho_i$
that has been derived in Ref.~\cite{Anders13NJP15} only on the basis of 
information-theoretic arguments. To see that note that $\rho_i$
can be seen as the thermal equilibrium associated to the Hamiltonian $H_i=-\beta^{-1} \ln \rho_i$,
and consider the case when the driving protocol is a sudden quench ($\tau \rightarrow 0$) of $H_i$ into $H_f$, 
in which case $\rho'= \rho_i$ and accordingly $ \langle Q \rangle=\Tr (\rho_f-\rho_i)H_f$.

In the following we will consider a thermodynamic process as illustrated above but of a special kind.
Namely we consider the case when $\tau \rightarrow 0$, i.e., the Hamiltonian is instantaneously quenched
from $H_i$ to $H_f$ at  time $t=0$. This means that in our specific case $U$ is the identity operator and that
$\rho'=\rho_i$. 
Further we focus on the special case when the initial Hamiltonian $H_i$ splits in the sum of two non-interacting 
sub-systems Hamiltonians
\begin{align}
H_i= H_A + H_B
\end{align}
and the quench protocol consists in turning on an interaction between them
\begin{align}
H_f= H_A + H_B + h
\end{align}
As a consequence of the irreversible thermalization step the initially uncorrelated state
\begin{align}
\rho_i = \frac{e^{-\beta H_A}}{Z_A} \otimes \frac{e^{-\beta H_B}}{Z_B} 
\label{eq:rhoi}
\end{align}
of the system will get correlated
\begin{align}
\rho_f = \frac{e^{-\beta (H_A+H_B+h)}}{Z_{f}}
\end{align}
where $Z_{x}= \Tr_x e^{-\beta H_{x}}$, with $x=A,B,f$ and $\Tr_x$ the trace over the respective Hilbert space.
We will investigate the work pdf, the dissipated work/heat and the correlation build-up, in this specific scenario.

%%%%%%%%%%%%%%%%%%%%%%%%%%%%%%%%%%%%%%%%%%%%%%%%%%%%
%%%%%%%%%%%%%%%%%%%%%%%%%%%%%%%%%%%%%%%%%%%%%%%%%%%%
\section{The Model}\label{S.Model}
%%%%%%%%%%%%%%%%%%%%%%%%%%%%%%%%%%%%%%%%%%%%%%%%%%%%
We consider two XX  chains of lengths $N_A$, $N_B$, which are joined %
 into a single XX chain of length $N=N_A+N_B$ \cite{Joshi13EPJB86}.
At times $t<0$, the system Hamiltonian is:
\begin{equation}
H_0 = H_A + H_B \,
\end{equation}
where
\begin{align}
H_A &= \frac{h}{2}\sum_{j=1}^{N_A} {\sigma_j ^z} - \frac{J}{4}\sum_{j=1} ^{N_A-1} {[\sigma_j ^x \sigma_{j+1} ^x + \sigma_j ^y \sigma_{j+1} ^y]}\, \\
H_B &= \frac{h}{2}\sum_{j=N_A+1}^{N} {\sigma_j ^z} - \frac{J}{4}\sum_{j=N_A+1} ^{N-1} {[\sigma_j ^x \sigma_{j+1} ^x + \sigma_j ^y \sigma_{j+1} ^y]}\,
\end{align}
with $\sigma_j^\alpha$, $j=1 \dots N$, $\alpha=x,y,z$, denoting the Pauli matrices of  the $j$-th spin.
At time $t=0$ an interaction between spin $N_A$ and spin $N_A+1$ is turned on,
such that the Hamiltonian is, for $t>0$:
\begin{align}
H_f &= \frac{h}{2}\sum_{j=1}^{N} {\sigma_j ^z} - \frac{J}{4}\sum_{j=1} ^{N-1} {[\sigma_j ^x \sigma_{j+1} ^x + \sigma_j ^y \sigma_{j+1} ^y]}\, .
\end{align}

The Hamiltonians $H_A,H_B,H_f$ all represent XX spin chains of different lengths $L$.
By means of Jordan-Wigner transformation
followed by a sine transform \cite{Lieb61AP16,Mikeska77ZPB26}
\begin{align}
 c_k &= \sqrt{\frac{2}{N+1}} \sum_{i=1}^{N} \sin \left( \frac{ki\pi}{N+1} \right)\left[\prod _{k=1}^{j-1}\sigma^z_k \sigma_j^{-}\right]
\label{eq:ck}
\end{align}
(with $\sigma_j^- = [\sigma^x_j - i \sigma^y_j$]/2) a chain of $L$ spins can be mapped on  $L$ non-interacting fermions, representing the 
normal modes of the chain \cite{Lieb61AP16,Joshi13EPJB86}
\begin{align}\label{E.HamFer}
H = \sum_{k=1}^{L} \varepsilon_{k}(L) {c}_{k}^{\dagger} {c}_{k},
\end{align}
where the one-body eigenenergies and eigenstates are given, respectively, by \cite{Son09PRA79}
\begin{align}
\varepsilon_{k}(L)&=h+ J\cos\frac{k \pi}{L+1}
\end{align} and $\ket{k}\equiv{c}_k^{\dagger}
\ket{0}$
with $\ket{0}$ being the fermion vacuum. The chain eigenenergies read then:
\begin{align}\label{E.HamFer}
E_\mathbf{n}=&\sum_{k=1}^{L} \varepsilon_{k}(L) n_k
\end{align}
where $\mathbf{n}=(n_i \dots n_L)$, denotes the occupation of each fermionic mode, $n_i=0,1$.

In the following and throughout the manuscript all energies are expressed in units of $J$ and we set 
$k_B$, Boltzmann's constant, equal to 1, accordingly temperature is measured in units of energy. 

Note that the three Hamiltonians $H_A,H_B,H_f$ have different 
lengths, hence different creation operators, different eigenvectors, and different single mode energies $\varepsilon_k(L)$.
To fix the notation, we shall denote the eigenvectors of the $H_A,H_B,H_f$ as $ |\mathbf{i}_A  \rangle, |\mathbf{i}_B \rangle$ and $|\mathbf{f}  \rangle$, respectively. The eigenstates $ |\mathbf{i}\rangle$ of $H_i$ are specified by the $N$ occupation numbers of the two sub-chains $|\mathbf{i}\rangle = |\mathbf{i}_A\rangle |\mathbf{i}_B\rangle$. The associated eigenenergies will be denoted by the symbols $E_\mathbf{i}$ and $E'_\mathbf{f}$.

As can be seen from the expression of the quasi-particle energy $\varepsilon_k\in\left[h-1,h+1\right]$, depending on the value of $h$, the energy of the $k$-mode can be either positive or negative, and this determines if, in the ground state, i.e., at $T=0$, that mode will be occupied or not by the corresponding fermion. At finite temperature, the $k$-mode has an occupation probability according to Fermi statistics. The last value of $k$ such that $\varepsilon_k\leq 0$ is called Fermi wave vector~\cite{Altland10}, and reads, for the discrete lattice model we are investigating, $k_F=\left[\frac{N+1}{\pi}\cos^{-1}h\right]$, where $[\cdot]$ denotes the integer part. As a consequence, the ground state of the model is
$\ket{GS}=\prod_{k\geq k_F}c_k^{\dagger}\ket{0}$.

It is well known that the model under scrutiny exhibits a quantum phase transition from a quasi-long range ordered state in the ferromagnetic phase to a disorder state in the paramagnetic phase~\cite{Niemeijer67P36}. This transition occurs, at $T=0$ and $N\rightarrow \infty$, for $h=1$. The paramagnetic state corresponds to
$\ket{GS}=\ket{0}$ ($\ket{GS}=\prod_{k=1}^N c_k^{\dagger}\ket{0}$) for $h\geq 1$ ($h\leq -1$). Expressed in the original spin language, it means that all the spins are aligned along the external magnetic field $h$, $\ket{GS}=\ket{1}^{\otimes N}$ ($\ket{GS}=\ket{0}^{\otimes N}$), where $\ket{1}$ ($\ket{0}$) represents the spin pointing down $\ket{\downarrow}$ (up $\ket{\uparrow}$) on each site. As a consequence, this state is referred to as \textit{factorized} state and the point $h=1$ as \textit{factorizing} field $h_f$~\cite{Kurmann82PA112,Amico06PRA74,Giampaolo09PRB79,Campbell13PRA88}. Since the sign of $h$ in the hamiltonian can be reversed by a unitary transformation embodying a rotation of all spins by $\pi$ around the $x$-axis~\cite{Safonov13PRA87}, in the following we will restrict, without loss of generality, to the interval $h\geq 0$. 
%%%%%%%%%%%%%%%%%%%%%%%%% FIGURE %%%%%%%%%%%
\begin{figure}
\includegraphics[width=\columnwidth]{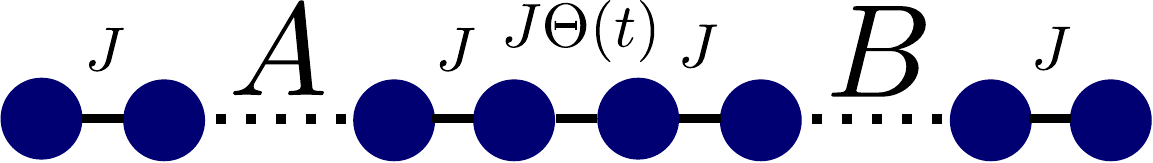}
 \caption{(Color online) Two spin chains $A$ and $B$ are joined at time $t=0$ by $J$.}
  \label{F.Model}
\end{figure}
%%%%%%%%%%%%%%%%%%%%%%%%% FIGURE %%%%%%%%%%%
For finite systems the factorization point does not correspond anymore to $h=1$, but it is given by $h_f= \cos\frac{\pi}{N+1}$ and, moreover, for open chains, factorization does not occur homogeneously throughout the whole chain because of the lack of translational invariance. In fact, the alignment along the magnetic field is less frustrated at the borders of the chain than in the bulk, where the competing presence of the other spins is more likely to contrast the alignment along the $z$-axis.

%%%%%%%%%%%%%%%%%%%%%%%%%%%%%%%%%%%%%%%%%%%%%%%%%%%%
%%%%%%%%%%%%%%%%%%%%%%%%%%%%%%%%%%%%%%%%%%%%%%%%%%%%
\section{Work statistics}
\label{work}
%%%%%%%%%%%%%%%%%%%%%%%%%%%%%%%%%%%%%%%%%%%%%%%%%%%%
According to the two-measurement scheme~\cite{Campisi11RMP83} the work probability distribution function associated to the instantaneous connection of the two chains is
\begin{align}
p(w)= \sum_{\mathbf{i},\mathbf{f}} \delta[w-(E'_\mathbf{f}-E_\mathbf{i})]P[{\mathbf{f}|\mathbf{i}}]e^{-\beta E_\mathbf{i}}/Z_i,
\label{eq:p(w)}
\end{align}
where $\delta(x)$ is Dirac delta function, $Z_i= \sum_{\mathbf{i}}e^{-\beta E_\mathbf{i}}$ is the partition function of the initial equilibrium, and $P[{\mathbf{f}|\mathbf{i}}]$ is the probability of finding the system in state $ | \mathbf{f} \rangle$ after the quench, provided it was in state $| \mathbf{i} \rangle$ before it. As the quench is instantaneous, we have $P[{\mathbf{f}|\mathbf{i}}]=|\langle \mathbf{i}|\mathbf{f}\rangle|^2$.
The overlaps $\langle \mathbf{i}|\mathbf{f}\rangle$ can be calculated analytically, see Appendix \ref{app:P[f|i]}. However, since there are $2^N\times 2^N$ of them, calculating them all is very lengthy (this notwithstanding the selection rule according to which the number of excitations before the quench must equal the number of excitations after the quench $\sum_k i_k=\sum_{l}f_l$). We have pursued the calculation for a maximum length $N$ of $10$ spins, corresponding to a transition matrix $P[{\mathbf{f}|\mathbf{i}}]$ of ca. $10^6$ elements.

\begin{figure}[]
		\begin{center}
		\includegraphics[width=\linewidth]{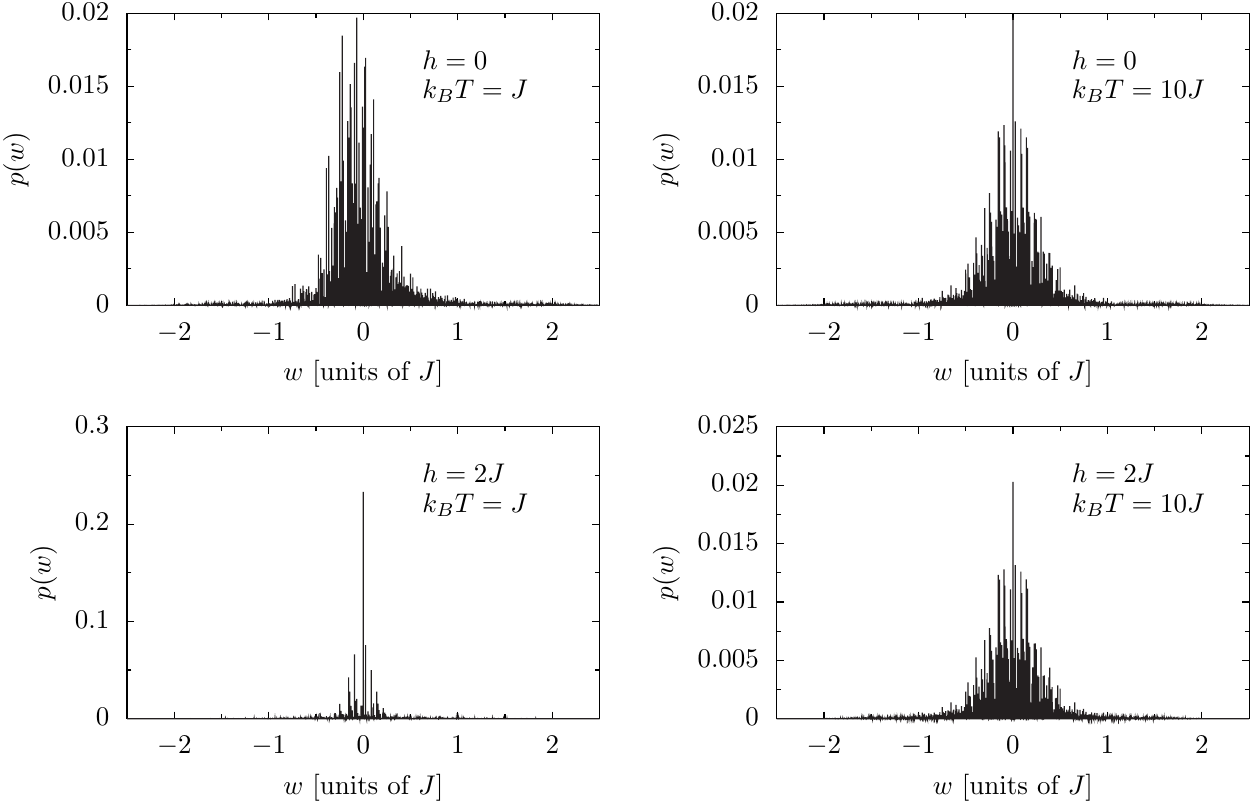}
		\caption{Work probability distribution function at various values of transverse magnetic field $h$ and temperature $T$}
		\label{fig:Fig2}
		\end{center}
\end{figure}

Figure \ref{fig:Fig2} reports the work pdf for $N_A=N_B=5$ at various values of $h$ and $k_B T$~\footnote{To be more precise the figure 
reports a histogram of the work probability. The bin size is chosen as $R_w/10^6$ where 
$R_w=\max_{\mathbf{f},\mathbf{i}}(E'_\mathbf{f}-E_\mathbf{i}) - \min_{\mathbf{f},\mathbf{i}}(E'_\mathbf{f}-E_\mathbf{i}) $ is the range of range within which the work spans.}. 
%%%%%%%%%%%%%%%%%%%%%%%%% FIGURE %%%%%%%%%%%
%%%%%%%%%%%%%%%%%%%%%%%%%%%%%%%%%%%%%%%%%%
Note how, with increasing the magnetic field, the width of the distribution decreases.
This is because the relative effect of the added interaction energy, of order $J$, becomes
smaller compared to the energy of the full chain. This effect becomes more and
more pronounced as the temperature decreases (see the bottom-left panel in Fig. \ref{fig:Fig2}),
leading the work pdf to  a single dirac delta centered around $w=0$ in the limit $T\rightarrow 0$,
where the only initial state entering the sum Eq. (\ref{eq:p(w)}) is a factorized state, having
non null overlap with itself only.

It is important to stress that, for this problem the average work done is null. From Eq. (\ref{eq:W-DF}) the average work
is
\begin{align}
\langle w \rangle &= \int dw p(w) w = \Tr [\rho' H_f-\rho_i H_i] 
\end{align}
hence in our case  where we have a sudden quench ($\rho'=\rho_f$)
\begin{align}
\langle w \rangle= -J\Tr [\rho_i (\sigma_{N_A} ^x \sigma_{N_A+1} ^x + \sigma_{N_A} ^y \sigma_{N_A+1} ^y])]/4=0
\end{align}
namely the average work is given by the initial $x$ and $y$ correlations between the last spin of the first 
chain and the first spin of the second chain. Since our initial state is a product state, those initial correlations
are null. Thus turning on the interaction has no cost in terms of energy \footnote{The situation is akin to that of quickly moving 
the piston that confines a gas as to half it volume. If the initial state of the gas corresponds to all its molecules being 
already confined in the final volume, no work is needed to move the piston.}. As a consequence, in the present case:
\begin{align}
W_\text{irr} = \Delta F\\
\langle Q \rangle = \Delta U
 \end{align}

The change in free energy can be expressed as \cite{Campisi10CP375}
\begin{align}\label{E.FreeEDiff}
 \Delta F&=-\frac{1}{\beta}\ln\frac{Z_f}{Z_i} \\
 &=-\frac{1}{\beta} \ln\frac{\prod_{q=1}^{N}[1+e^{-\beta \varepsilon_q(N) }]}{\prod_{q=1}^{N_A}[1+e^{-\beta \varepsilon_q(N_A) }]\prod_{q=1}^{N_B}[1+e^{-\beta \varepsilon_q(N_B) }]}\nonumber
\end{align}

%%%%%%%%%%%%%%%%%%%%%%%%%%%%%%%%%%%%%%%%%%%%%%%%%%%%
%%%%%%%%%%%%%%%%%%%%%%%%%%%%%%%%%%%%%%%%%%%%%%%%%%%%
\section{Irreversible Work}
\label{sec:IrrW}
%%%%%%%%%%%%%%%%%%%%%%%%%%%%%%%%%%%%%%%%%%%%%%%%%%%%

%%%%%%%%%%%%%%%%%%%%%%%%%%%%%%%%%%%%%%%%%%%%%%%%%%%%
%%%%%%%%%%%%%%%%%%%%%%%%%%%%%%%%%%%%%%%%%%%%%%%%%%%%
\subsection{Varying the magnetic field}
\label{S.quenchvsh}
%%%%%%%%%%%%%%%%%%%%%%%%%%%%%%%%%%%%%%%%%%%%%%%%%%%%
In this Section we evaluate the irreversible work $W_\text{irr}$  at different initial magnetic fields and relate its features to the many-body physics of the model, especially to the presence of level crossings and quantum phase transitions.

\begin{figure}
 \centering
   \includegraphics[width=\columnwidth]{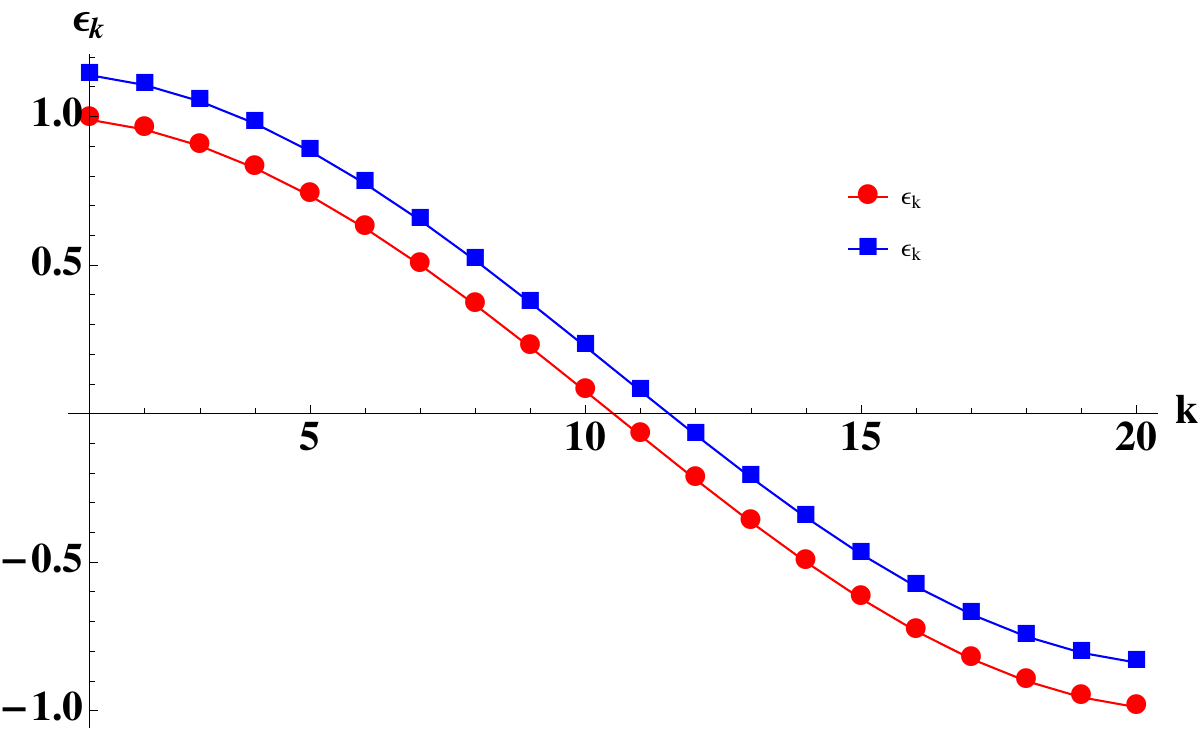}\\
\caption{(Color online) Single mode energy spectrum for a spin chain of length $N=20$, at  $h=0$ (red) and $h=0.15$ (blue). 
At $h=0$, 10 modes have positive energy, while at $h=0.15$, there are 11 modes at positive energy. The quench $h=0 \rightarrow 0.15$ realizes one level crossing.}
  \label{F.QPT1}
\end{figure}

\begin{figure}
 \centering
   \includegraphics[width=\columnwidth]{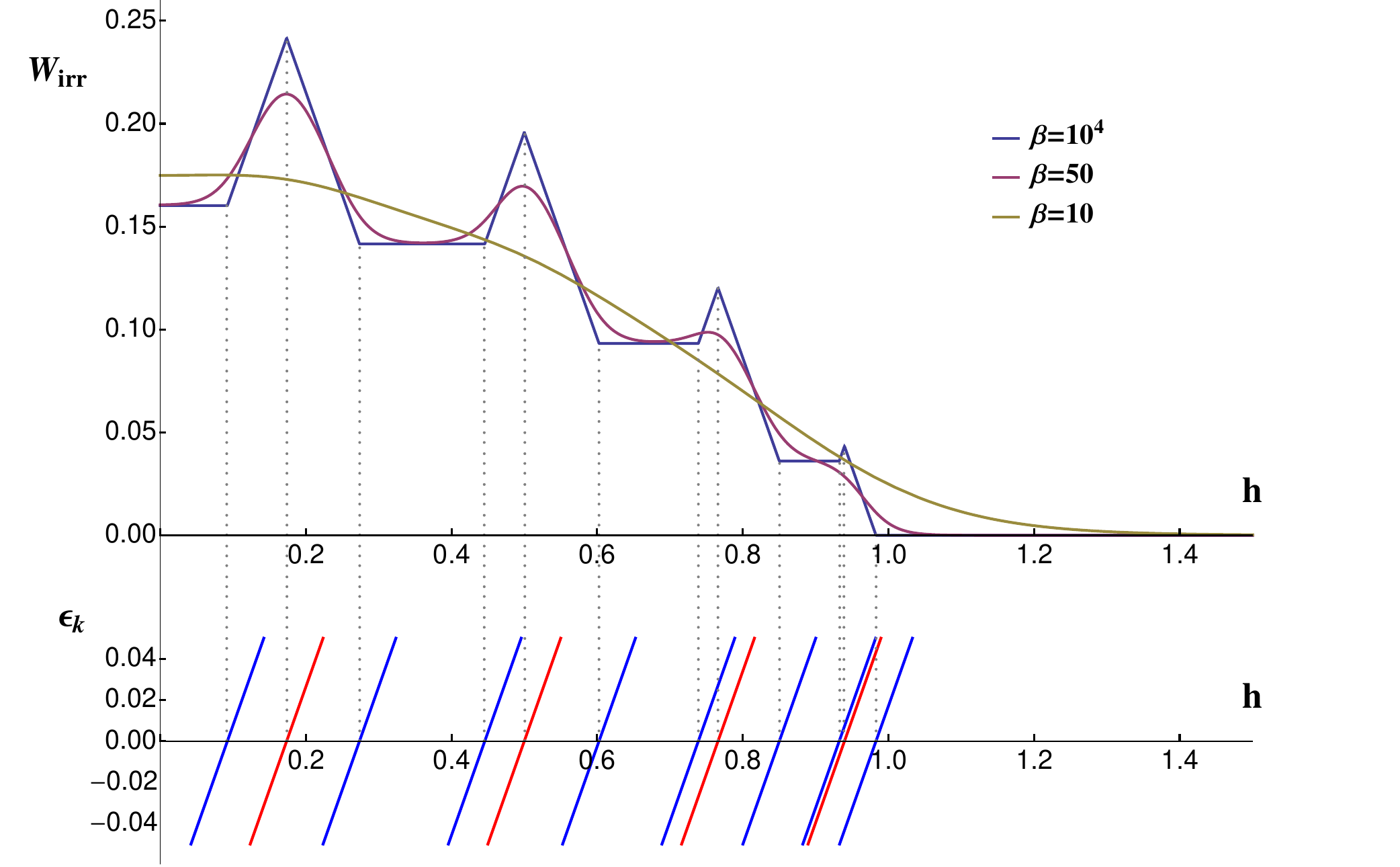}
\caption{(Color online) Irreversible work $W_\text{irr}$ done by joining two spin chains made of $N_A=N_B=8$ spins at inverse temperature $\beta=10^4, 500, 10$. On the lower part of the figure the pre-quench energies of the $k=1,2,...,4$-modes, each 4-fold degenerate, is shown (red lines) together with the corresponding post-quench energies (blue lines), close to the crossing points. Vertical lines are to shown the one-to-one correspondence between the crossings and the non-analiticities of $W_\text{irr}$ at $T = 0$.}
  \label{F.QPT3}
\end{figure}

The model under scrutiny  exhibits, at $T=0$, $N$ level crossings as $h$ is varied \cite{Son09PRA79,Campbell13NJP15}, each occurring at $h_k(N)=- \cos{\frac{k \pi}{N+1}}$, when the energy of the $k$-fermion $\varepsilon_k=h+\cos{\frac{k \pi}{N+1}}$ changes sign. Accordingly the ground state  $\ket{GS}=\prod_{k\geq k_f}c_k^{\dagger}\ket{0}$ changes its structure at those values of $h$.
In the interval between consecutive $h_k$'s, on the other hand, the ground state remains unchanged. In  Fig.~\ref{F.QPT1} an instance of such a level crossing is illustrated when the global magnetic field is varied. The presence of these level crossings is signaled by a peak in the the irreversible work plotted as a function of $h$ for small global quenches $\Delta h$ \cite{Francica} in a way similar to that reported in \cite{Dorner12PRL109,Fusco14arxiv} for the Ising model.
%%%%%%%%%%%%%%%%%%%%%%%%% FIGURE %%%%%%%%%%%
%%%%%%%%%%%%%%%%%%%%%%%%% FIGURE %%%%%%%%%%%
However the present situation is different. We do not have here a global quench of the magnetic field $h$, but rather
a local quench of the interaction $J$ between the two chains. The turning on of this interaction can change
the structure of the ground state in a similar way as it happens in a global $h$ quench. 
This fact is illustrated in Fig. \ref{F.QPT3}. In the bottom graph we plot the single mode energies of the pre- and post- quench 
Hamiltonians $H_i$ and $H_f$ with $N_A=N_B $. Before the quench each single mode is four-fold degenerate (a factor two comes from the $k \leftrightarrow -k$ symmetry and a second factor two comes from the left right symmetry: $N_A=N_B$).
The left-right symmetry degeneracy is lifted when the central bond is turned on, so that each level splits into two levels, one staying above and one staying below the original one. The structure of the ground state is affected then when one of these has a different sign than the original un-splitted mode. This is reflected in the triangular peaks al very low temperature in Fig. \ref{F.QPT3} top.
These peaks are smoothed out as the temperature increases as expected \cite{Dorner12PRL109,Fusco14arxiv,Francica}.
%%%%%%%%%%%%%%%%%%%%%%%%% FIGURE %%%%%%%%%%%
%%%%%%%%%%%%%%%%%%%%%%%%% FIGURE %%%%%%%%%%%

%%%%%%%%%%%%%%%%%%%%%%%%%%%%%%%%%%%%%%%%%%%%%%%%%%%%
%%%%%%%%%%%%%%%%%%%%%%%%%%%%%%%%%%%%%%%%%%%%%%%%%%%%
\subsection{Quenches from the factorized phase into the critical one}
\label{sS.equalNhfact}
%%%%%%%%%%%%%%%%%%%%%%%%%%%%%%%%%%%%%%%%%%%%%%%%%%%%

\begin{figure}
 \centering
      \includegraphics[width=\columnwidth]{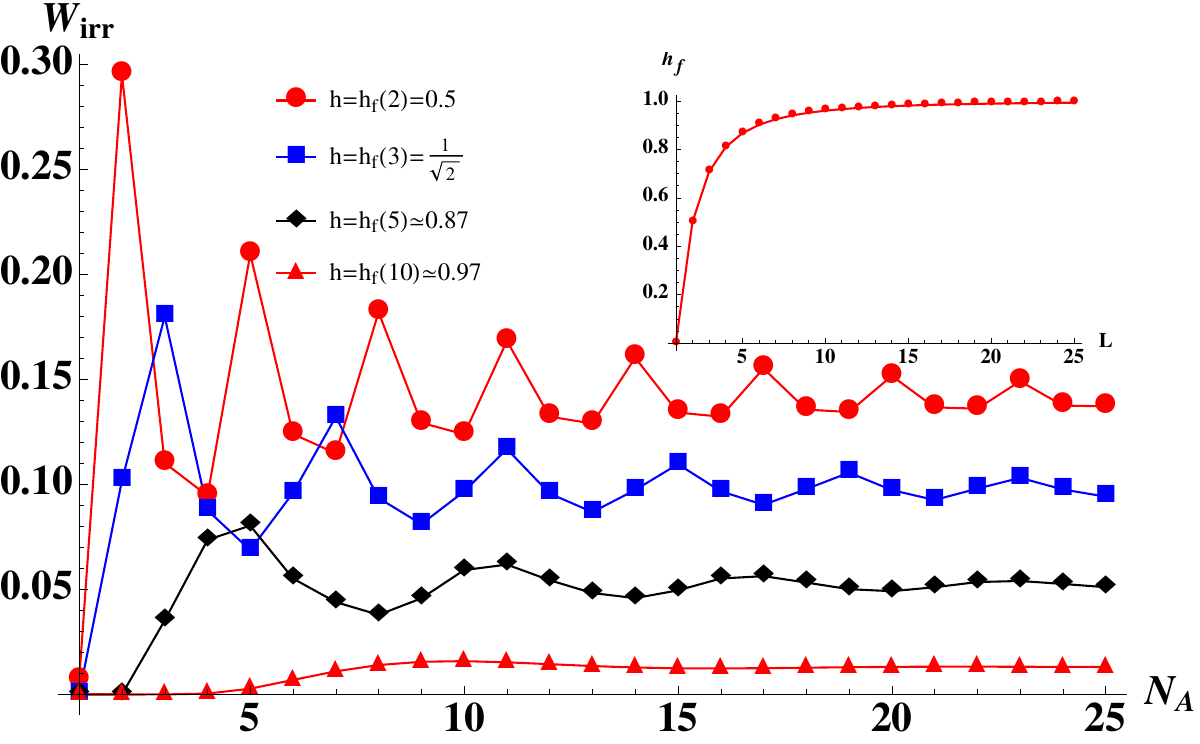}
       \includegraphics[width=\columnwidth]{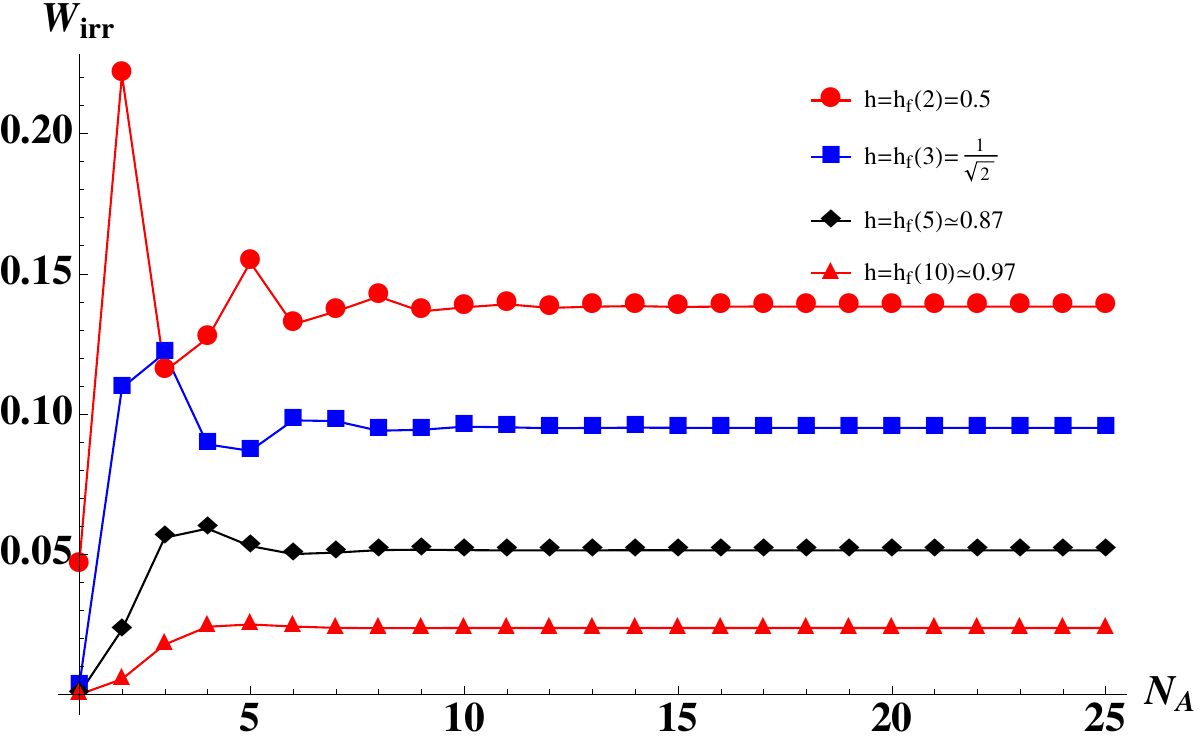}
\caption{(Color online)Inset: Factorizing magnetic field $h_f$ as a function of the length of the chain $N_A$. Notice that the factorizing field increases with the number of spins, thus there are values of the magnetic field $h$ that result in a factorizing magnetic field for each of the pre-quenched systems $N_A=N_B$, but not for the post quench system made of $N_A+N_B$ spins.
Top panel: irreversible work $W_\text{irr}$ as function of $N_A$ at different magnetic fields and $\beta=10^2$. The magnetic fields are chosen in such a way they coincide with a factorizing field at a certain $N_A$. These are marked by the first peak of $W_\text{irr}$.
Bottom panel: same as top panel but for a higher temperature $\beta = 15$. The oscillatory behavior is washed out but the first maximum survives.}
  \label{F.hfat}
\end{figure}

In the previous section~\ref{S.quenchvsh}, we have investigated what happens when turning on the interaction between 
the two chains corresponds to a level crossing. In this Section we investigate the behavior of the irreversible work as the last level-crossing, namely the one that separates the
factorized (i.e. paramagnetic) and un-factorized (i.e. ferromagnetic) phases at zero temperature. 
As already stated in Sec.~\ref{S.Model},  at $T=0$ and $N\rightarrow \infty$, the factorizing field is $h_f=1$. Nevertheless,  for finite chains, the factorizing field $h_f$ is a function of $N$. Fig.~\ref{F.hfat} shows the dependence of the factorizing field $h_f$ on the chain length $L$. The important aspect to note here is 
that $h_f$ is a monotonically increasing function of $L$. Therefore a value of $h$ which is a factorizing one for each of the two sub-chains might not be a factorizing field for the whole, longer chain.

Fig.~\ref{F.hfat} shows the irreversible work $W_\text{irr}$ as a function of $N_A(=N_B)$ at specific values of the magnetic fields, $h=0,5,\frac{1}{\sqrt{2}},0.87,0.97$, which, respectively, are the factorizing fields for $N_A=2,3,5,10$.
We observe that the global maximum of $W_\text{irr}$ occurs at the value of $N_A$ for which the field is factorizing. This is
because the turning on of the interaction between the two chains leads into the un-factorized phase of the longer chain, accordingly. At higher temperature this effect is smoothed out as expected, see Fig.~\ref{F.hfat}, bottom panel.
%%%%%%%%%%%%%%%%%%%%%%%%% FIGURE %%%%%%%%%%%
%%%%%%%%%%%%%%%%%%%%%%%%% FIGURE %%%%%%%%%%%

%%%%%%%%%%%%%%%%%%%%%%%%%%%%%%%%%%%%%%%%%%%%%%%%%%%%
\subsection{Role of the inhomogeneity of the spin alignment}\label{sS.difN}
%%%%%%%%%%%%%%%%%%%%%%%%%%%%%%%%%%%%%%%%%%%%%%%%%%%%
Because of boundary effects (due to the lack of translational invariance) and finite temperature effects, the alignment of all the spins along the $z$-direction, is not achieved uniformly along the chain even above the factorizing field. Close to the boundaries, the spins tend to align with the magnetic field more easily. This in-homogeneity is reflected in the behavior of $W_\text{irr}$ as function of $N_A$, and is more marked for values of $h$ close to the factorizing field.

\begin{figure}
 \centering
       \includegraphics[width=\columnwidth]{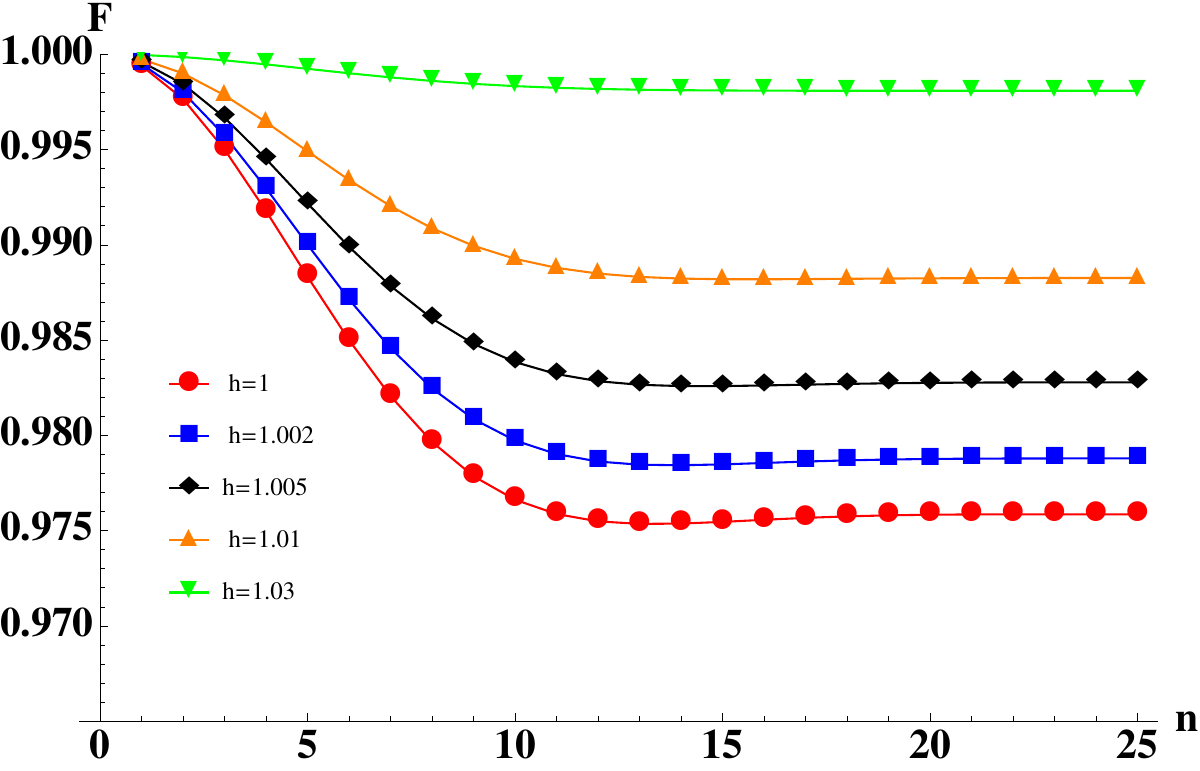}\\
        \includegraphics[width=\columnwidth]{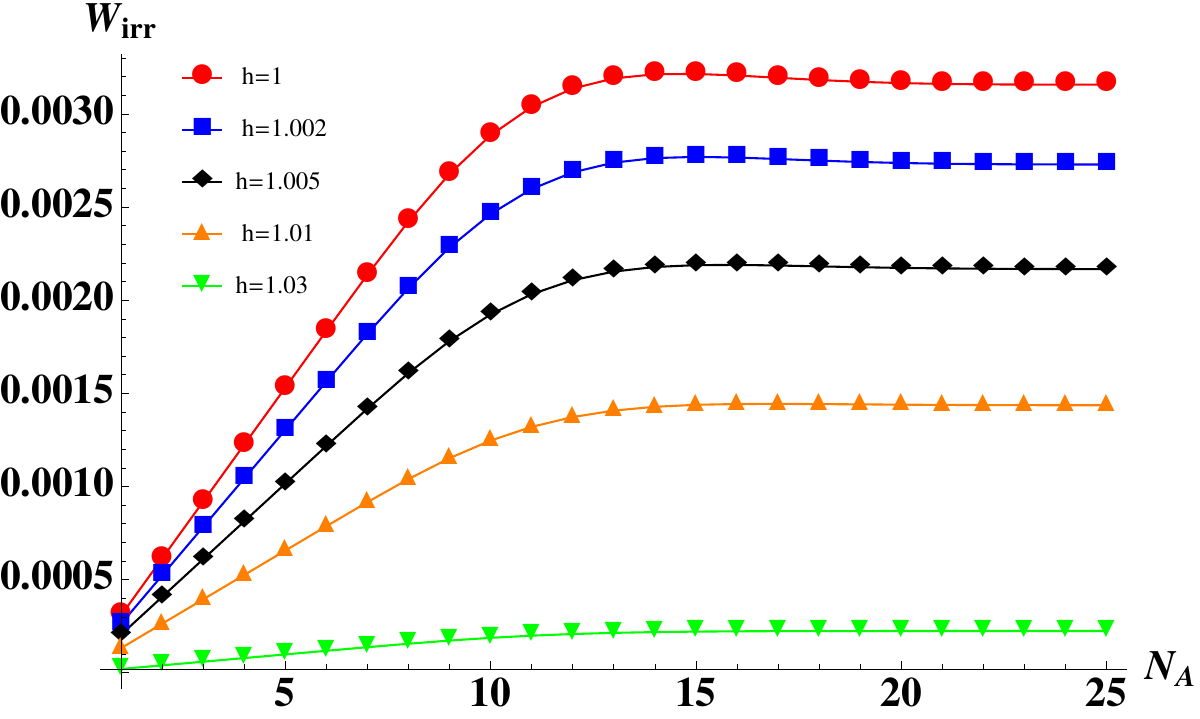}
       \caption{(Color online)  Top panel: Fidelity between the spin located at $n$ in an $N=50$ open chain and a fully aligned state along $z$ at different magnetic fields and $\beta=100$. Close to the borders the magnetic field induces a stronger factorization, measured by a higher fidelity.  Bottom panel: Irreversible work $W_\text{irr}$  done by joining two chains of lenght $N_A+N_B=50$,  as a function of $N_A$ for values of $h\gtrsim 1$. The irreversible work increases with increasing $N_A$ until it reaches a bulk value.}
  \label{F.DifNhvar}
\end{figure}

The degree of alignment along the chain can be conveniently quantified by the fidelity $F\!\left(\tilde{\rho}_n,\rho_n\right)=\valmed{0\vert\rho_n\vert 0}$ \cite{Josza94JMO41} between the reduced density matrix  $\rho_n$ of the $n$'th spin and the state it would have if completely aligned with the magnetic field, i.e., $\tilde{\rho}_n=\ket{0}\!\!\bra{0}$. Using the expression 
$\rho_n=\text{diag}[1+\valmed{\sigma_n^z}, 1-\valmed{\sigma_n^z}]  /{2}$, the fidelity amounts to  $F=(1+\valmed{\sigma_n^z})/{2}$.  
Fig.~\ref{F.DifNhvar} top panel,  shows $F$ as a function of the spin location for a chain of $N=50$ spins at $\beta = 100$
Note that, as anticipated, close to the boundaries the factorization effect of the magnetic field $h$ is stronger and that the boundary/finite temperature effects are less marked as $h$ increases beyond the factorizing value $h_f(50)  \lesssim 1$.

This phenomenon is reflected in the behavior of $W_\text{irr}$ as a function of  $N_A$ for fixed $N$.
When the joining occurs close to the boundaries, small $N_A$, where the alignment in the $z$ direction is more complete, the quench has a lower effect on the irreversible work, as compared to quenches occurring in the bulk. 
This is because when quenching close to the boundary both initial and final states are well aligned, hence are ``closer'' on to the other, and their Kullback-Leibler divergence (proportional to $W_\text{irr}$, see Eq.~\ref{eq:KL}) is small. In the bulk however the pre-quench state is appreciably more aligned than the post-quench state, hence more ``distant'' to it.
Note that a plateaux value is reached when
the junction occurs deep enough into the bulk. Note also how the bulk value of the irreversible work diminishes when the 
magnetic field is increased. This is because with increasing field more alignment is present in both the initial and final states 
$\rho_i$ and $\rho_f$, which as a consequence differ less one from the other.
%%%%%%%%%%%%%%%%%%%%%%%%% FIGURE %%%%%%%%%%%
%%%%%%%%%%%%%%%%%%%%%%%%% FIGURE %%%%%%%%%%%

%%%%%%%%%%%%%%%%%%%%%%%%%%%%%%%%%%%%%%%%%%%%%%%%%%%%
\section{Build up of correlations}
\label{sS.equalN}
%%%%%%%%%%%%%%%%%%%%%%%%%%%%%%%%%%%%%%%%%%%%%%%%%%%%
As the system is let thermalize after the quench, it reaches a state $\rho_f$ in which the two sub-chains 
are correlated. Since the initial state $\rho_i$ was red a product state, see Eq. (\ref{eq:rhoi}), the thermalization process leads 
to a build up of correlation. At the same time, as discussed in Sec. \ref{tools}, heat is exchanged with the bath.
Here we investigate both the behavior of the irreversible work $W_\text{irr}$ (that is the negative irreversible heat $Q_\text{irr}$ exchanged with the bath in the thermalization step), and of the correlation build-up, as function of $N_A=N_B=L$.

As quantifiers of correlations we focus on the concurrence $C$~\cite{Wootters98PRL80} and the mutual information $MI$ of the 
spins located at $n= N_A$ and $n=N_{A}+1$. While the concurrence is a measure of genuine quantum correlations, the mutual information contains both classical and quantum correlations.

\begin{figure}
 \centering
   \includegraphics[width=\columnwidth]{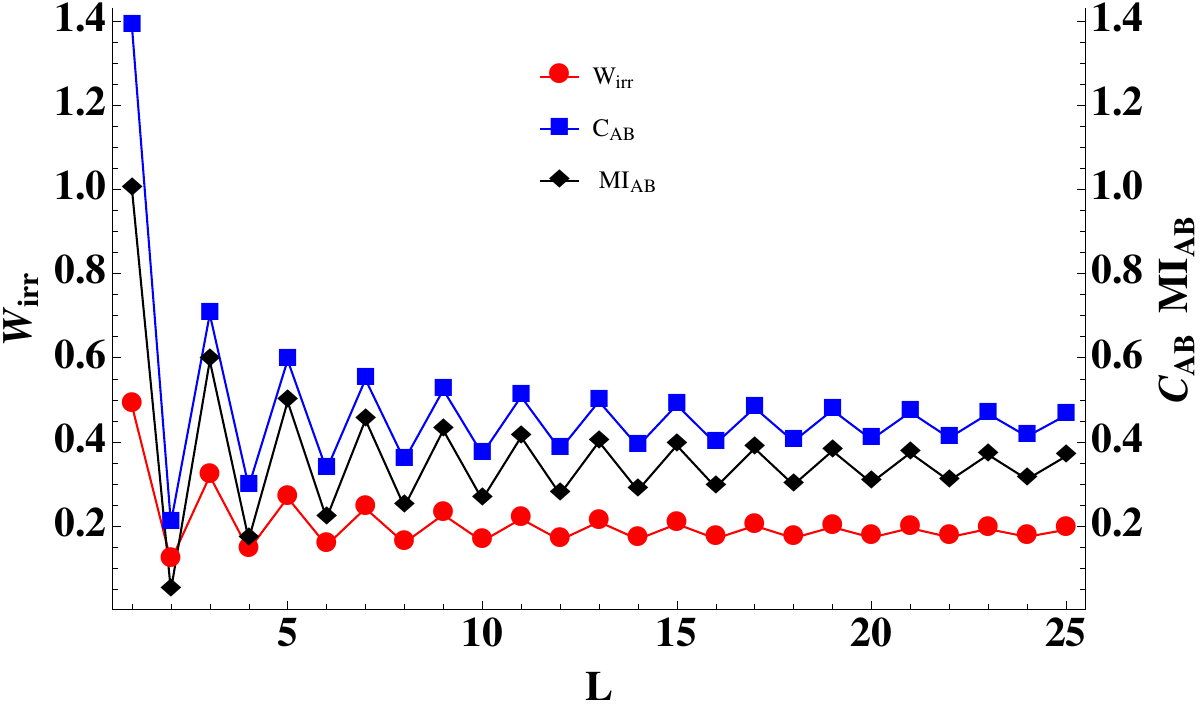}\\
      \includegraphics[width=\columnwidth]{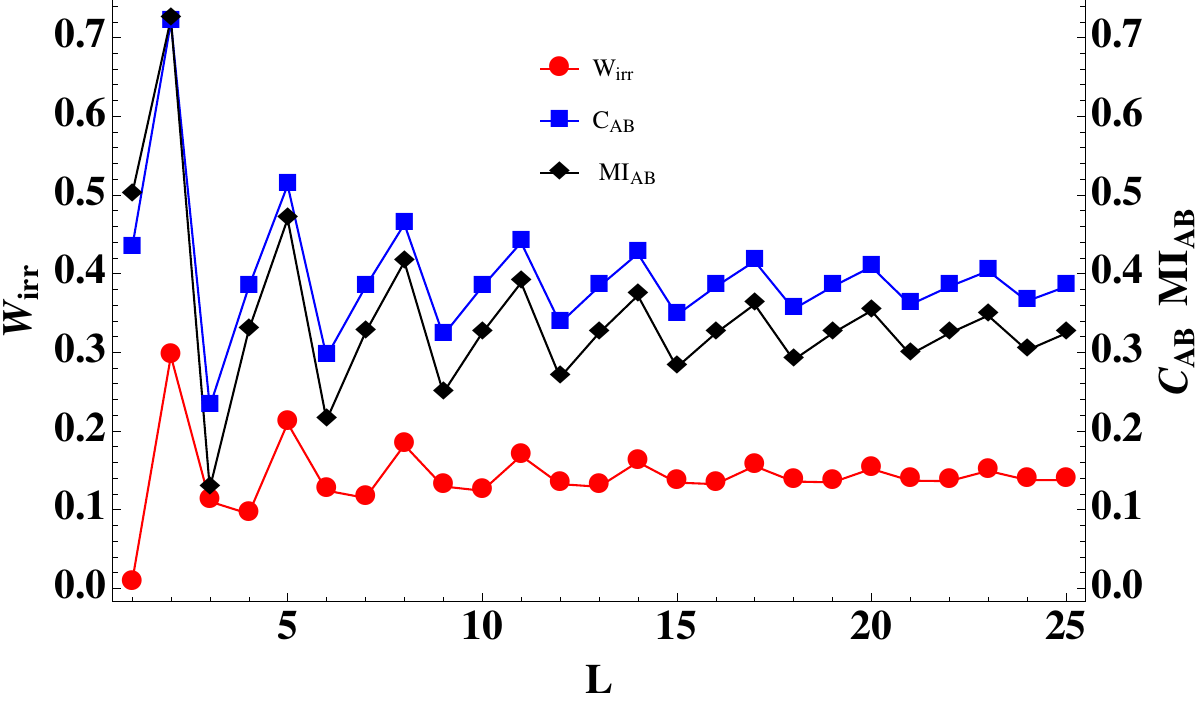}\\
       \includegraphics[width=\columnwidth]{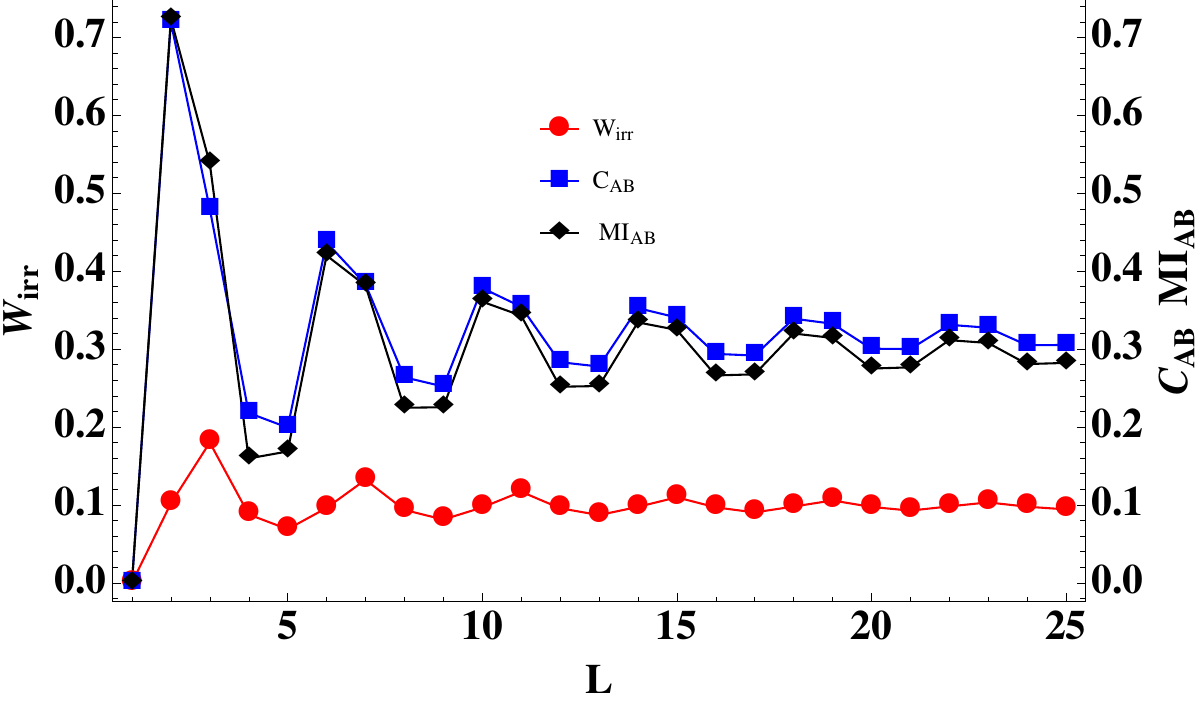}
\caption{(Color online) Irreversible work $W_\text{irr}$, concurrence $C$ and mutual information $MI$ between spin $N_A$ and $N_A+1$ in the final thermal equilibrium state of two chains at different lengths $N_A=N_B=L$ of the connected chains. The figures are reported for $h{=}0, \frac{1}{2}, \frac{1}{\sqrt{2}} $ (top-down) and $\beta{=}10^2$.}
  \label{F.DifLenhvarh0}
\end{figure}

Using the notation $\rho_a$, $\rho_b$, $\rho_{ab}$ for the reduced density matrices of the $N^{\rm th}_A$ spin, the $N^{\rm th}_B$ spin
and the two of them, respectively, the concurrence is given by  $C= \max\left[0, 2|\rho^{ab}_{12}|-\sqrt{\rho^{00}_{ab}\rho^{33}_{ab}}\right]/2$~\cite{Amico04PRA69,Apollaro13PRA88}, where $\rho^{ij}_{ab}$ is the $(i,j)$-element of $\rho_{ab}$ in the computational basis.
The mutual information  is given by  $MI = S(\rho_a)+S(\rho_b)-S(\rho_{ab})$, where $S(\cdot)$ is the usual von Neumann entropy. 
Fig.~\ref{F.DifLenhvarh0} shows the behavior of $W_\text{irr}$, $C$ and $MI$ for different values of $h$. An oscillatory behaviour of all these quantities by varying $L$ is clearly visible with the period of the oscillation given by  $p=\pi/(\cos^{-1}h)$.

The oscillations of $MI$ and $C$ can be traced back to boundary effects. In fact, the presence of the borders is responsible for stronger spin-spin correlations, which enter the expression of both $MI$ and $C$, the closer the spins are to the edges.
This boundary effect has been investigated both in chains made of a finite number of particles~\cite{Son09PRA79} and in infinite systems close to impurities which effectively break the chain~\cite{Apollaro08PRA77}. The fact that $W_\text{irr}$ presents the same oscillatory behaviour suggests that the main contribution in this case comes from the correlations between spin $N_A$ and spin $N_A+1$.
This latter consideration can be also supported by considering that the irreversible work is the  Kullback-Leibler divergence between $\rho^i$ and $\rho^f$, and noting that far from the joining region the two density matrices are not so different, whereas, the main contribution arises from the neighborhood of the connected borders.
%%%%%%%%%%%%%%%%%%%%%%%%% FIGURE %%%%%%%%%%%
%%%%%%%%%%%%%%%%%%%%%%%%% FIGURE %%%%%%%%%%%

%%%%%%%%%%%%%%%%%%%%%%%%%%%%%%%%%%%%%%%%%%%%%%%%%%%%
%%%%%%%%%%%%%%%%%%%%%%%%%%%%%%%%%%%%%%%%%%%%%%%%%%%%
\section{Conclusions}\label{sec:conclusions}
%%%%%%%%%%%%%%%%%%%%%%%%%%%%%%%%%%%%%%%%%%%%%%%%%%%%
We have studied the influences of quantum many-body effects, such as criticality and the existence of factorisation fields, in the thermodynamic cost of establishing a bonding link between two independent quantum spin chains ruled by an XX model. The behavior of both reversible and irreversible work produced in such process has been interpreted by establishing an explicit link to the phenomenology of such quantities to the properties of the spectrum of the system.We have included the effects of an external magnetic field, as well as inhomogeneities, thus providing an extensive analysis of the various key parameters in the spin model addressed in our study. Our work contributes to the ongoing effort directed towards the understanding of the many-body implications for non-equilibrium thermodynamics.  

%%%%%%%%%%%%%%%%%%%%%%%%%%%%%%%%%%%%%%%%%%%%%%%%%%%%
\appendix

\section{Calculation of $P[{\mathbf f}|{\mathbf i}]$}
\label{app:P[f|i]}
In order to calculate the transition probabilities $P[{\mathbf f}|{\mathbf i}]=|\langle \mathbf{f}|\mathbf{i}\rangle|^2$ we write \cite{Joshi13EPJB86}
\begin{align}
\langle \mathbf{f}|\mathbf{i}\rangle = \sum_{\mathbf s} \langle \mathbf{f}|\mathbf{s}\rangle \langle \mathbf{s}|\mathbf{i}\rangle
\label{eq:<f|i>}
\end{align}
where $|\mathbf{s}\rangle= |s_1 \dots s_{N_A},s_{N_A+1} \dots s_N\rangle $, $s_i=\pm 1$, is the eigenbasis of 
$\sigma_1^z \otimes \sigma_2^z \otimes \dots \otimes \sigma_N^z$. Recall that $|\mathbf i\rangle$ is an eigenvector of
$H_A + H_B$, as such it is of the form $|\mathbf i_A\rangle |\mathbf i_B\rangle$ with $|\mathbf i_{A(B)}\rangle$ the eigenvector of $H_{A(B)}$. Likewise, $|\mathbf{s}\rangle$, can be written as $|\mathbf{s}\rangle= |\mathbf s_A\rangle |\mathbf s_B\rangle$.
So that $\langle \mathbf{s}|\mathbf{i}\rangle= \langle \mathbf{s}_A|\mathbf{i}_A\rangle \langle \mathbf{s}_B|\mathbf{i}_B\rangle$.
We proceed then to write an explicit formula for $\langle \mathbf{f}|\mathbf{s}\rangle$. Using the fermionic rule 
\begin{align} 
c_k^\dagger |...\,  f_k \,...  \rangle &= (1-f_k)(-1)^{\sum_{i=1}^{k-1}f_i} | ...\,  f_k+1 \,...  \rangle  \label{eq:bk3}
\end{align}
and the explicit expression for the rising operators, Eq. (\ref{eq:ck}),
one finds, after a careful inspection, the general formula:
\begin{align}
\langle \mathbf{f}|\mathbf{s}\rangle = \delta_{r,t} \prod_{i=1}^N s_i^{G(i,\mathbf{q})} \sum_\sigma S(k_1 q_{\sigma(r)}) \dots S(k_r q_{\sigma(1)})\mathcal P(\sigma) \label{eq:<f|s>}
\end{align}
where the symbols appearing in Eq. (\ref{eq:<f|s>}) have the following meaning. Given $|\mathbf f \rangle$, the ordered labels $k_1 < \dots< k_r$ indicate which modes are excited in the state $|\mathbf f \rangle$. For example if $|\mathbf f \rangle = |0,1,1\rangle$, there are two excitations, $r=2$, and $k_1=2,k_2=3$, saying that the second and third modes are excited. Similarly $q_1 < \dots <q_t$ say which spins are up in the state $|\mathbf s\rangle $. Note that $\langle \mathbf{f}|\mathbf{s}\rangle$ is automatically  zero if $t \neq r$, hence the Kronecker symbol $ \delta_{r,t}$. The symbol $\sigma$ stands here for a permutation of  $\{1,2, \dots r \}$, for example 
$\{2,1, \dots r \}$, in which case $\sigma(1)=2, \sigma(2)=1, \dots \sigma(r)=r$. $\mathcal P(\sigma)$ is the parity of the permutation 
$\sigma$ and the sum is over all the $r!$ possible permutations. $G(i,\mathbf q)$ is the number of elements in the vector $\mathbf{q}= (q_1 \dots q_t)$ which are larger than $i$, and $S(x)=\sqrt{2/(N+1)}\sin[x\pi/(N+1)]$. The general formula can be used for any chain length, so, it can be used to calculate $\langle \mathbf{s}_A|\mathbf{i}_A\rangle$ and $ \langle \mathbf{s}_B|\mathbf{i}_B\rangle$ as well, hence $\langle \mathbf{s}|\mathbf{i}\rangle$. This combined with $\langle \mathbf{f}|\mathbf{s}\rangle$ gives
$\langle \mathbf{f}|\mathbf{i}\rangle$ as from Eq. (\ref{eq:<f|i>}), and by squaring, finally gives $P[{\mathbf f}|{\mathbf i}]=|\langle \mathbf{f}|\mathbf{i}\rangle|^2$

\acknowledgements
T.J.G.A. was supported by the European Commission, the European Social Fund and the Region Calabria through the program POR Calabria FSE 2007-2013ÐAsse IV Capitale Umano-Obiettivo Operativo M2. M.C. was supported by a Marie Curie Intra European Fellowship within the 7th European Community Framework Programme through the project NeQuFlux grant n. 623085. M.P. thanks the UK EPSRC for a Career Acceleration Fellowship and a grant awarded under the ``New Directions for Research Leaders'' initiative (EP/G004579/1), the John Templeton Foundation (grant 43467), and the EU Collaborative Project TherMiQ (grant egreement 618074). This work is partially supported by the COST Action MP1209.


\begin{thebibliography}{50}
\expandafter\ifx\csname natexlab\endcsname\relax\def\natexlab#1{#1}\fi
\expandafter\ifx\csname bibnamefont\endcsname\relax
  \def\bibnamefont#1{#1}\fi
\expandafter\ifx\csname bibfnamefont\endcsname\relax
  \def\bibfnamefont#1{#1}\fi
\expandafter\ifx\csname citenamefont\endcsname\relax
  \def\citenamefont#1{#1}\fi
\expandafter\ifx\csname url\endcsname\relax
  \def\url#1{\texttt{#1}}\fi
\expandafter\ifx\csname urlprefix\endcsname\relax\def\urlprefix{URL }\fi
\providecommand{\bibinfo}[2]{#2}
\providecommand{\eprint}[2][]{\url{#2}}

\bibitem[{\citenamefont{Esposito et~al.}(2009)\citenamefont{Esposito, Harbola,
  and Mukamel}}]{Esposito09RMP81}
\bibinfo{author}{\bibfnamefont{M.}~\bibnamefont{Esposito}},
  \bibinfo{author}{\bibfnamefont{U.}~\bibnamefont{Harbola}}, \bibnamefont{and}
  \bibinfo{author}{\bibfnamefont{S.}~\bibnamefont{Mukamel}},
  \bibinfo{journal}{Rev. Mod. Phys.} \textbf{\bibinfo{volume}{81}},
  \bibinfo{pages}{1665} (\bibinfo{year}{2009}).

\bibitem[{\citenamefont{Campisi et~al.}(2011)\citenamefont{Campisi, H{\"a}nggi,
  and Talkner}}]{Campisi11RMP83}
\bibinfo{author}{\bibfnamefont{M.}~\bibnamefont{Campisi}},
  \bibinfo{author}{\bibfnamefont{P.}~\bibnamefont{H{\"a}nggi}},
  \bibnamefont{and} \bibinfo{author}{\bibfnamefont{P.}~\bibnamefont{Talkner}},
  \bibinfo{journal}{Rev. Mod. Phys.} \textbf{\bibinfo{volume}{83}},
  \bibinfo{pages}{771} (\bibinfo{year}{2011}), \bibinfo{note}{\emph{ibid.}
  \textbf{83} 1653 (2011).}

\bibitem[{\citenamefont{Tasaki}(2000)}]{Tasaki00arxiv}
\bibinfo{author}{\bibfnamefont{H.}~\bibnamefont{Tasaki}},
  \bibinfo{journal}{arXiv:cond-mat/0009244}  (\bibinfo{year}{2000}).

\bibitem[{\citenamefont{Kurchan}(2000)}]{Kurchan00arxiv}
\bibinfo{author}{\bibfnamefont{J.}~\bibnamefont{Kurchan}},
  \bibinfo{journal}{arXiv:cond-mat/0007360v2}  (\bibinfo{year}{2000}).

\bibitem[{\citenamefont{Talkner et~al.}(2007)\citenamefont{Talkner, Lutz, and
  H{\"a}nggi}}]{Talkner07PRE75}
\bibinfo{author}{\bibfnamefont{P.}~\bibnamefont{Talkner}},
  \bibinfo{author}{\bibfnamefont{E.}~\bibnamefont{Lutz}}, \bibnamefont{and}
  \bibinfo{author}{\bibfnamefont{P.}~\bibnamefont{H{\"a}nggi}},
  \bibinfo{journal}{Phys. Rev. E} \textbf{\bibinfo{volume}{75}},
  \bibinfo{pages}{050102} (\bibinfo{year}{2007}).

\bibitem[{\citenamefont{{Campisi} et~al.}(2009)\citenamefont{{Campisi},
  {Talkner}, and {H{\"a}nggi}}}]{Campisi09PRL102}
\bibinfo{author}{\bibfnamefont{M.}~\bibnamefont{{Campisi}}},
  \bibinfo{author}{\bibfnamefont{P.}~\bibnamefont{{Talkner}}},
  \bibnamefont{and}
  \bibinfo{author}{\bibfnamefont{P.}~\bibnamefont{{H{\"a}nggi}}},
  \bibinfo{journal}{Phys. Rev. Lett.} \textbf{\bibinfo{volume}{102}},
  \bibinfo{pages}{210401} (\bibinfo{year}{2009}).

\bibitem[{\citenamefont{Talkner et~al.}(2009)\citenamefont{Talkner, Campisi,
  and H{\"a}nggi}}]{Talkner09JSM09}
\bibinfo{author}{\bibfnamefont{P.}~\bibnamefont{Talkner}},
  \bibinfo{author}{\bibfnamefont{M.}~\bibnamefont{Campisi}}, \bibnamefont{and}
  \bibinfo{author}{\bibfnamefont{P.}~\bibnamefont{H{\"a}nggi}},
  \bibinfo{journal}{J. Stat. Mech.: Theory Exp.} p. \bibinfo{pages}{P02025}
  (\bibinfo{year}{2009}).

\bibitem[{\citenamefont{Pekola et~al.}(2012)\citenamefont{Pekola, Solinas,
  Shnirman, and Averin}}]{Pekola12arxiv}
\bibinfo{author}{\bibfnamefont{J.~P.} \bibnamefont{Pekola}},
  \bibinfo{author}{\bibfnamefont{P.}~\bibnamefont{Solinas}},
  \bibinfo{author}{\bibfnamefont{A.}~\bibnamefont{Shnirman}}, \bibnamefont{and}
  \bibinfo{author}{\bibfnamefont{D.~V.} \bibnamefont{Averin}},
  \bibinfo{journal}{arXiv:1212.5808}  (\bibinfo{year}{2012}).

\bibitem[{\citenamefont{Dorner et~al.}(2013)\citenamefont{Dorner, Clark,
  Heaney, Fazio, Goold, and Vedral}}]{Dorner13PRL110}
\bibinfo{author}{\bibfnamefont{R.}~\bibnamefont{Dorner}},
  \bibinfo{author}{\bibfnamefont{S.~R.} \bibnamefont{Clark}},
  \bibinfo{author}{\bibfnamefont{L.}~\bibnamefont{Heaney}},
  \bibinfo{author}{\bibfnamefont{R.}~\bibnamefont{Fazio}},
  \bibinfo{author}{\bibfnamefont{J.}~\bibnamefont{Goold}}, \bibnamefont{and}
  \bibinfo{author}{\bibfnamefont{V.}~\bibnamefont{Vedral}},
  \bibinfo{journal}{Phys. Rev. Lett.} \textbf{\bibinfo{volume}{110}},
  \bibinfo{pages}{230601} (\bibinfo{year}{2013}).

\bibitem[{\citenamefont{Mazzola et~al.}(2013)\citenamefont{Mazzola, De~Chiara,
  and Paternostro}}]{Mazzola13PRL110}
\bibinfo{author}{\bibfnamefont{L.}~\bibnamefont{Mazzola}},
  \bibinfo{author}{\bibfnamefont{G.}~\bibnamefont{De~Chiara}},
  \bibnamefont{and}
  \bibinfo{author}{\bibfnamefont{M.}~\bibnamefont{Paternostro}},
  \bibinfo{journal}{Phys. Rev. Lett.} \textbf{\bibinfo{volume}{110}},
  \bibinfo{pages}{230602} (\bibinfo{year}{2013}).

\bibitem[{\citenamefont{Campisi et~al.}(2013)\citenamefont{Campisi, Blattmann,
  Kohler, Zueco, and H{\"a}nggi}}]{Campisi13NJP15}
\bibinfo{author}{\bibfnamefont{M.}~\bibnamefont{Campisi}},
  \bibinfo{author}{\bibfnamefont{R.}~\bibnamefont{Blattmann}},
  \bibinfo{author}{\bibfnamefont{S.}~\bibnamefont{Kohler}},
  \bibinfo{author}{\bibfnamefont{D.}~\bibnamefont{Zueco}}, \bibnamefont{and}
  \bibinfo{author}{\bibfnamefont{P.}~\bibnamefont{H{\"a}nggi}},
  \bibinfo{journal}{New Journal of Physics} \textbf{\bibinfo{volume}{15}},
  \bibinfo{pages}{105028} (\bibinfo{year}{2013}).

\bibitem[{\citenamefont{Mazzola et~al.}(2014)\citenamefont{Mazzola, De~Chiara,
  and Paternostro}}]{Mazzola14arxiv}
\bibinfo{author}{\bibfnamefont{L.}~\bibnamefont{Mazzola}},
  \bibinfo{author}{\bibfnamefont{G.}~\bibnamefont{De~Chiara}},
  \bibnamefont{and}
  \bibinfo{author}{\bibfnamefont{M.}~\bibnamefont{Paternostro}},
  \bibinfo{journal}{arXiv:1401.0566}  (\bibinfo{year}{2014}).

\bibitem[{\citenamefont{{Liphardt} et~al.}(2002)\citenamefont{{Liphardt},
  {Dumont}, {Smith}, {Tinoco}, and {Bustamante}}}]{Liphardt02SCIENCE296}
\bibinfo{author}{\bibfnamefont{J.}~\bibnamefont{{Liphardt}}},
  \bibinfo{author}{\bibfnamefont{S.}~\bibnamefont{{Dumont}}},
  \bibinfo{author}{\bibfnamefont{S.~B.} \bibnamefont{{Smith}}},
  \bibinfo{author}{\bibfnamefont{I.}~\bibnamefont{{Tinoco}}}, \bibnamefont{and}
  \bibinfo{author}{\bibfnamefont{C.}~\bibnamefont{{Bustamante}}},
  \bibinfo{journal}{Science} \textbf{\bibinfo{volume}{296}},
  \bibinfo{pages}{1832} (\bibinfo{year}{2002}).

\bibitem[{\citenamefont{{Collin} et~al.}(2005)\citenamefont{{Collin}, {Ritort},
  {Jarzynski}, {Smith}, {Tinoco}, and {Bustamante}}}]{Collin05NAT437}
\bibinfo{author}{\bibfnamefont{D.}~\bibnamefont{{Collin}}},
  \bibinfo{author}{\bibfnamefont{F.}~\bibnamefont{{Ritort}}},
  \bibinfo{author}{\bibfnamefont{C.}~\bibnamefont{{Jarzynski}}},
  \bibinfo{author}{\bibfnamefont{S.~B.} \bibnamefont{{Smith}}},
  \bibinfo{author}{\bibfnamefont{I.}~\bibnamefont{{Tinoco}}}, \bibnamefont{and}
  \bibinfo{author}{\bibfnamefont{C.}~\bibnamefont{{Bustamante}}},
  \bibinfo{journal}{Nature} \textbf{\bibinfo{volume}{437}},
  \bibinfo{pages}{231} (\bibinfo{year}{2005}).

\bibitem[{\citenamefont{Douarche et~al.}(2005)\citenamefont{Douarche,
  Ciliberto, Petrosyan, and Rabbiosi}}]{Douarche05EPL70}
\bibinfo{author}{\bibfnamefont{F.}~\bibnamefont{Douarche}},
  \bibinfo{author}{\bibfnamefont{S.}~\bibnamefont{Ciliberto}},
  \bibinfo{author}{\bibfnamefont{A.}~\bibnamefont{Petrosyan}},
  \bibnamefont{and} \bibinfo{author}{\bibfnamefont{I.}~\bibnamefont{Rabbiosi}},
  \bibinfo{journal}{Europhys. Lett.} \textbf{\bibinfo{volume}{70}},
  \bibinfo{pages}{593} (\bibinfo{year}{2005}).

\bibitem[{\citenamefont{Toyabe et~al.}(2010)\citenamefont{Toyabe, Sagawa, Ueda,
  Muneyuki, and Sano}}]{Toyabe10NP6}
\bibinfo{author}{\bibfnamefont{S.}~\bibnamefont{Toyabe}},
  \bibinfo{author}{\bibfnamefont{T.}~\bibnamefont{Sagawa}},
  \bibinfo{author}{\bibfnamefont{M.}~\bibnamefont{Ueda}},
  \bibinfo{author}{\bibfnamefont{E.}~\bibnamefont{Muneyuki}}, \bibnamefont{and}
  \bibinfo{author}{\bibfnamefont{M.}~\bibnamefont{Sano}},
  \bibinfo{journal}{Nature Physics} \textbf{\bibinfo{volume}{6}},
  \bibinfo{pages}{988} (\bibinfo{year}{2010}).

\bibitem[{\citenamefont{Saira et~al.}(2012)\citenamefont{Saira, Yoon, Tanttu,
  M{\"o}tt{\"o}nen, Averin, and Pekola}}]{Saira12PRL109}
\bibinfo{author}{\bibfnamefont{O.-P.} \bibnamefont{Saira}},
  \bibinfo{author}{\bibfnamefont{Y.}~\bibnamefont{Yoon}},
  \bibinfo{author}{\bibfnamefont{T.}~\bibnamefont{Tanttu}},
  \bibinfo{author}{\bibfnamefont{M.}~\bibnamefont{M{\"o}tt{\"o}nen}},
  \bibinfo{author}{\bibfnamefont{D.~V.} \bibnamefont{Averin}},
  \bibnamefont{and} \bibinfo{author}{\bibfnamefont{J.~P.}
  \bibnamefont{Pekola}}, \bibinfo{journal}{Phys. Rev. Lett.}
  \textbf{\bibinfo{volume}{109}}, \bibinfo{pages}{180601}
  (\bibinfo{year}{2012}).

\bibitem[{\citenamefont{Batalh\~{a}o et~al.}(2013)\citenamefont{Batalh\~{a}o,
  Souza, Mazzola, Auccaise, Oliveira, Goold, De~Chiara, Paternostro, and
  Serra}}]{Batalhao13arxiv}
\bibinfo{author}{\bibfnamefont{T.}~\bibnamefont{Batalh\~{a}o}},
  \bibinfo{author}{\bibfnamefont{A.~M.} \bibnamefont{Souza}},
  \bibinfo{author}{\bibfnamefont{L.}~\bibnamefont{Mazzola}},
  \bibinfo{author}{\bibfnamefont{R.}~\bibnamefont{Auccaise}},
  \bibinfo{author}{\bibfnamefont{I.~S.} \bibnamefont{Oliveira}},
  \bibinfo{author}{\bibfnamefont{J.}~\bibnamefont{Goold}},
  \bibinfo{author}{\bibfnamefont{G.}~\bibnamefont{De~Chiara}},
  \bibinfo{author}{\bibfnamefont{M.}~\bibnamefont{Paternostro}},
  \bibnamefont{and} \bibinfo{author}{\bibfnamefont{R.~M.} \bibnamefont{Serra}},
  \bibinfo{journal}{arXiv:1308.3241}  (\bibinfo{year}{2013}).

\bibitem[{\citenamefont{Crooks}(1999)}]{Crooks99PRE60}
\bibinfo{author}{\bibfnamefont{G.~E.} \bibnamefont{Crooks}},
  \bibinfo{journal}{Phys. Rev. E} \textbf{\bibinfo{volume}{60}},
  \bibinfo{pages}{2721} (\bibinfo{year}{1999}).

\bibitem[{\citenamefont{Jarzynski}(1997)}]{Jarzynski97PRL78}
\bibinfo{author}{\bibfnamefont{C.}~\bibnamefont{Jarzynski}},
  \bibinfo{journal}{Phys. Rev. Lett.} \textbf{\bibinfo{volume}{78}},
  \bibinfo{pages}{2690} (\bibinfo{year}{1997}).

\bibitem[{\citenamefont{Silva}(2008)}]{Silva08PRL101}
\bibinfo{author}{\bibfnamefont{A.}~\bibnamefont{Silva}},
  \bibinfo{journal}{Phys. Rev. Lett.} \textbf{\bibinfo{volume}{101}},
  \bibinfo{pages}{120603} (\bibinfo{year}{2008}).

\bibitem[{\citenamefont{Dorner et~al.}(2012)\citenamefont{Dorner, Goold,
  Cormick, Paternostro, and Vedral}}]{Dorner12PRL109}
\bibinfo{author}{\bibfnamefont{R.}~\bibnamefont{Dorner}},
  \bibinfo{author}{\bibfnamefont{J.}~\bibnamefont{Goold}},
  \bibinfo{author}{\bibfnamefont{C.}~\bibnamefont{Cormick}},
  \bibinfo{author}{\bibfnamefont{M.}~\bibnamefont{Paternostro}},
  \bibnamefont{and} \bibinfo{author}{\bibfnamefont{V.}~\bibnamefont{Vedral}},
  \bibinfo{journal}{Phys. Rev. Lett.} \textbf{\bibinfo{volume}{109}},
  \bibinfo{pages}{160601} (\bibinfo{year}{2012}).

\bibitem[{\citenamefont{Smacchia and Silva}(2013)}]{Smacchia13PRE88}
\bibinfo{author}{\bibfnamefont{P.}~\bibnamefont{Smacchia}} \bibnamefont{and}
  \bibinfo{author}{\bibfnamefont{A.}~\bibnamefont{Silva}},
  \bibinfo{journal}{Phys. Rev. E} \textbf{\bibinfo{volume}{88}},
  \bibinfo{pages}{042109} (\bibinfo{year}{2013}).

\bibitem[{\citenamefont{Sindona et~al.}(2013)\citenamefont{Sindona, Goold,
  Lo~Gullo, Lorenzo, and Plastina}}]{Sindona13PRL111}
\bibinfo{author}{\bibfnamefont{A.}~\bibnamefont{Sindona}},
  \bibinfo{author}{\bibfnamefont{J.}~\bibnamefont{Goold}},
  \bibinfo{author}{\bibfnamefont{N.}~\bibnamefont{Lo~Gullo}},
  \bibinfo{author}{\bibfnamefont{S.}~\bibnamefont{Lorenzo}}, \bibnamefont{and}
  \bibinfo{author}{\bibfnamefont{F.}~\bibnamefont{Plastina}},
  \bibinfo{journal}{Phys. Rev. Lett.} \textbf{\bibinfo{volume}{111}},
  \bibinfo{pages}{165303} (\bibinfo{year}{2013}).

\bibitem[{\citenamefont{Mascarenhas et~al.}(2013)\citenamefont{Mascarenhas,
  Braganca, Dorner, Santos, Vedral, Modi, and Goold}}]{Mascarenhas13arxiv}
\bibinfo{author}{\bibfnamefont{E.}~\bibnamefont{Mascarenhas}},
  \bibinfo{author}{\bibfnamefont{H.}~\bibnamefont{Braganca}},
  \bibinfo{author}{\bibfnamefont{R.}~\bibnamefont{Dorner}},
  \bibinfo{author}{\bibfnamefont{M.~F.} \bibnamefont{Santos}},
  \bibinfo{author}{\bibfnamefont{V.}~\bibnamefont{Vedral}},
  \bibinfo{author}{\bibfnamefont{K.}~\bibnamefont{Modi}}, \bibnamefont{and}
  \bibinfo{author}{\bibfnamefont{J.}~\bibnamefont{Goold}},
  \bibinfo{journal}{arXiv:1307.5544}  (\bibinfo{year}{2013}).

\bibitem[{\citenamefont{Sindona et~al.}(2014)\citenamefont{Sindona, Goold,
  Lo~Gullo, and Plastina}}]{Sindona14NJP16}
\bibinfo{author}{\bibfnamefont{A.}~\bibnamefont{Sindona}},
  \bibinfo{author}{\bibfnamefont{J.}~\bibnamefont{Goold}},
  \bibinfo{author}{\bibfnamefont{N.}~\bibnamefont{Lo~Gullo}}, \bibnamefont{and}
  \bibinfo{author}{\bibfnamefont{F.}~\bibnamefont{Plastina}},
  \bibinfo{journal}{New J. Phys.} \textbf{\bibinfo{volume}{16}},
  \bibinfo{pages}{045013} (\bibinfo{year}{2014}).

\bibitem[{\citenamefont{Carlisle et~al.}(2014)\citenamefont{Carlisle, Mazzola,
  Campisi, Goold, Semi{\~a}o, Ferraro, Plastina, Vedral, De~Chiara, and
  Paternostro}}]{Carlisle14arxiv}
\bibinfo{author}{\bibfnamefont{A.}~\bibnamefont{Carlisle}},
  \bibinfo{author}{\bibfnamefont{L.}~\bibnamefont{Mazzola}},
  \bibinfo{author}{\bibfnamefont{M.}~\bibnamefont{Campisi}},
  \bibinfo{author}{\bibfnamefont{J.}~\bibnamefont{Goold}},
  \bibinfo{author}{\bibfnamefont{F.~L.} \bibnamefont{Semi{\~a}o}},
  \bibinfo{author}{\bibfnamefont{A.}~\bibnamefont{Ferraro}},
  \bibinfo{author}{\bibfnamefont{F.}~\bibnamefont{Plastina}},
  \bibinfo{author}{\bibfnamefont{V.}~\bibnamefont{Vedral}},
  \bibinfo{author}{\bibfnamefont{G.}~\bibnamefont{De~Chiara}},
  \bibnamefont{and}
  \bibinfo{author}{\bibfnamefont{M.}~\bibnamefont{Paternostro}},
  \bibinfo{journal}{arXiv:1403.0629}  (\bibinfo{year}{2014}).

\bibitem[{\citenamefont{{Fusco} et~al.}(2014)\citenamefont{{Fusco}, {Pigeon},
  {Apollaro}, {Xuereb}, {Mazzola}, {Campisi}, {Ferraro}, {Paternostro}, and {De
  Chiara}}}]{Fusco14arxiv}
\bibinfo{author}{\bibfnamefont{L.}~\bibnamefont{{Fusco}}},
  \bibinfo{author}{\bibfnamefont{S.}~\bibnamefont{{Pigeon}}},
  \bibinfo{author}{\bibfnamefont{T.~J.~G.} \bibnamefont{{Apollaro}}},
  \bibinfo{author}{\bibfnamefont{A.}~\bibnamefont{{Xuereb}}},
  \bibinfo{author}{\bibfnamefont{L.}~\bibnamefont{{Mazzola}}},
  \bibinfo{author}{\bibfnamefont{M.}~\bibnamefont{{Campisi}}},
  \bibinfo{author}{\bibfnamefont{A.}~\bibnamefont{{Ferraro}}},
  \bibinfo{author}{\bibfnamefont{M.}~\bibnamefont{{Paternostro}}},
  \bibnamefont{and} \bibinfo{author}{\bibfnamefont{G.}~\bibnamefont{{De
  Chiara}}}, \bibinfo{journal}{arXiv:1404.3150}  (\bibinfo{year}{2014}).

\bibitem[{\citenamefont{Crooks}(2008)}]{Crooks08JSM08}
\bibinfo{author}{\bibfnamefont{G.~E.} \bibnamefont{Crooks}},
  \bibinfo{journal}{J. Stat. Mech.: Theory Exp.} p. \bibinfo{pages}{P10023}
  (\bibinfo{year}{2008}).

\bibitem[{\citenamefont{Anders and Giovannetti}(2013)}]{Anders13NJP15}
\bibinfo{author}{\bibfnamefont{J.}~\bibnamefont{Anders}} \bibnamefont{and}
  \bibinfo{author}{\bibfnamefont{V.}~\bibnamefont{Giovannetti}},
  \bibinfo{journal}{New J. Phys.} \textbf{\bibinfo{volume}{15}},
  \bibinfo{pages}{033022} (\bibinfo{year}{2013}).

\bibitem[{\citenamefont{Vaikuntanathan and
  Jarzynski}(2009)}]{Vaikuntanathan09EPL87}
\bibinfo{author}{\bibfnamefont{S.}~\bibnamefont{Vaikuntanathan}}
  \bibnamefont{and}
  \bibinfo{author}{\bibfnamefont{C.}~\bibnamefont{Jarzynski}},
  \bibinfo{journal}{EPL} \textbf{\bibinfo{volume}{87}}, \bibinfo{pages}{60005}
  (\bibinfo{year}{2009}).

\bibitem[{\citenamefont{Joshi and Campisi}(2013)}]{Joshi13EPJB86}
\bibinfo{author}{\bibfnamefont{D.~G.} \bibnamefont{Joshi}} \bibnamefont{and}
  \bibinfo{author}{\bibfnamefont{M.}~\bibnamefont{Campisi}},
  \bibinfo{journal}{Eur. Phys. J. B} \textbf{\bibinfo{volume}{86}},
  \bibinfo{pages}{157} (\bibinfo{year}{2013}).

\bibitem[{\citenamefont{Lieb et~al.}(1961)\citenamefont{Lieb, Schultz, and
  Mattis}}]{Lieb61AP16}
\bibinfo{author}{\bibfnamefont{E.}~\bibnamefont{Lieb}},
  \bibinfo{author}{\bibfnamefont{T.}~\bibnamefont{Schultz}}, \bibnamefont{and}
  \bibinfo{author}{\bibfnamefont{D.}~\bibnamefont{Mattis}},
  \bibinfo{journal}{Ann. Phys.} \textbf{\bibinfo{volume}{16}},
  \bibinfo{pages}{407} (\bibinfo{year}{1961}).

\bibitem[{\citenamefont{Mikeska and Pesch}(1977)}]{Mikeska77ZPB26}
\bibinfo{author}{\bibfnamefont{H.~J.} \bibnamefont{Mikeska}} \bibnamefont{and}
  \bibinfo{author}{\bibfnamefont{W.}~\bibnamefont{Pesch}}, \bibinfo{journal}{Z.
  Phys. B} \textbf{\bibinfo{volume}{26}}, \bibinfo{pages}{351}
  (\bibinfo{year}{1977}).

\bibitem[{\citenamefont{Son et~al.}(2009)\citenamefont{Son, Amico, Plastina,
  and Vedral}}]{Son09PRA79}
\bibinfo{author}{\bibfnamefont{W.}~\bibnamefont{Son}},
  \bibinfo{author}{\bibfnamefont{L.}~\bibnamefont{Amico}},
  \bibinfo{author}{\bibfnamefont{F.}~\bibnamefont{Plastina}}, \bibnamefont{and}
  \bibinfo{author}{\bibfnamefont{V.}~\bibnamefont{Vedral}},
  \bibinfo{journal}{Phys. Rev. A} \textbf{\bibinfo{volume}{79}},
  \bibinfo{pages}{022302} (\bibinfo{year}{2009}).

\bibitem[{\citenamefont{Altland and Simons}(2010)}]{Altland10}
\bibinfo{author}{\bibfnamefont{A.}~\bibnamefont{Altland}} \bibnamefont{and}
  \bibinfo{author}{\bibfnamefont{B.}~\bibnamefont{Simons}},
  \emph{\bibinfo{title}{Consended Matter Field Theory}} (\bibinfo{year}{2010}).

\bibitem[{\citenamefont{Niemeijer}(1967)}]{Niemeijer67P36}
\bibinfo{author}{\bibfnamefont{T.}~\bibnamefont{Niemeijer}},
  \bibinfo{journal}{Physica} \textbf{\bibinfo{volume}{36}},
  \bibinfo{pages}{377} (\bibinfo{year}{1967}).

\bibitem[{\citenamefont{Kurmann et~al.}(1982)\citenamefont{Kurmann, Thomas, and
  M{\"u}ller}}]{Kurmann82PA112}
\bibinfo{author}{\bibfnamefont{J.}~\bibnamefont{Kurmann}},
  \bibinfo{author}{\bibfnamefont{H.}~\bibnamefont{Thomas}}, \bibnamefont{and}
  \bibinfo{author}{\bibfnamefont{G.}~\bibnamefont{M{\"u}ller}},
  \bibinfo{journal}{Physica A: Statistical Mechanics and its Applications}
  \textbf{\bibinfo{volume}{112}}, \bibinfo{pages}{235 } (\bibinfo{year}{1982}).

\bibitem[{\citenamefont{Amico et~al.}(2006)\citenamefont{Amico, Baroni, Fubini,
  Patan\`e, Tognetti, and Verrucchi}}]{Amico06PRA74}
\bibinfo{author}{\bibfnamefont{L.}~\bibnamefont{Amico}},
  \bibinfo{author}{\bibfnamefont{F.}~\bibnamefont{Baroni}},
  \bibinfo{author}{\bibfnamefont{A.}~\bibnamefont{Fubini}},
  \bibinfo{author}{\bibfnamefont{D.}~\bibnamefont{Patan\`e}},
  \bibinfo{author}{\bibfnamefont{V.}~\bibnamefont{Tognetti}}, \bibnamefont{and}
  \bibinfo{author}{\bibfnamefont{P.}~\bibnamefont{Verrucchi}},
  \bibinfo{journal}{Phys. Rev. A} \textbf{\bibinfo{volume}{74}},
  \bibinfo{pages}{022322} (\bibinfo{year}{2006}).

\bibitem[{\citenamefont{Giampaolo et~al.}(2009)\citenamefont{Giampaolo, Adesso,
  and Illuminati}}]{Giampaolo09PRB79}
\bibinfo{author}{\bibfnamefont{S.~M.} \bibnamefont{Giampaolo}},
  \bibinfo{author}{\bibfnamefont{G.}~\bibnamefont{Adesso}}, \bibnamefont{and}
  \bibinfo{author}{\bibfnamefont{F.}~\bibnamefont{Illuminati}},
  \bibinfo{journal}{Phys. Rev. B} \textbf{\bibinfo{volume}{79}},
  \bibinfo{pages}{224434} (\bibinfo{year}{2009}).

\bibitem[{\citenamefont{Campbell
  et~al.}(2013{\natexlab{a}})\citenamefont{Campbell, Richens, Gullo, and
  Busch}}]{Campbell13PRA88}
\bibinfo{author}{\bibfnamefont{S.}~\bibnamefont{Campbell}},
  \bibinfo{author}{\bibfnamefont{J.}~\bibnamefont{Richens}},
  \bibinfo{author}{\bibfnamefont{N.~L.} \bibnamefont{Gullo}}, \bibnamefont{and}
  \bibinfo{author}{\bibfnamefont{T.}~\bibnamefont{Busch}},
  \bibinfo{journal}{Phys. Rev. A} \textbf{\bibinfo{volume}{88}},
  \bibinfo{pages}{062305} (\bibinfo{year}{2013}{\natexlab{a}}).

\bibitem[{\citenamefont{Safonov and Lychkovskiy}(2013)}]{Safonov13PRA87}
\bibinfo{author}{\bibfnamefont{E.}~\bibnamefont{Safonov}} \bibnamefont{and}
  \bibinfo{author}{\bibfnamefont{O.}~\bibnamefont{Lychkovskiy}},
  \bibinfo{journal}{Phys. Rev. A} \textbf{\bibinfo{volume}{87}},
  \bibinfo{pages}{042105} (\bibinfo{year}{2013}).

\bibitem[{\citenamefont{Campisi et~al.}(2010)\citenamefont{Campisi, Zueco, and
  Talkner}}]{Campisi10CP375}
\bibinfo{author}{\bibfnamefont{M.}~\bibnamefont{Campisi}},
  \bibinfo{author}{\bibfnamefont{D.}~\bibnamefont{Zueco}}, \bibnamefont{and}
  \bibinfo{author}{\bibfnamefont{P.}~\bibnamefont{Talkner}},
  \bibinfo{journal}{Chem. Phys.} \textbf{\bibinfo{volume}{375}},
  \bibinfo{pages}{187} (\bibinfo{year}{2010}).

\bibitem[{\citenamefont{Campbell
  et~al.}(2013{\natexlab{b}})\citenamefont{Campbell, Mazzola, De~Chiara,
  Apollaro, Plastina, Busch, and Paternostro}}]{Campbell13NJP15}
\bibinfo{author}{\bibfnamefont{S.}~\bibnamefont{Campbell}},
  \bibinfo{author}{\bibfnamefont{L.}~\bibnamefont{Mazzola}},
  \bibinfo{author}{\bibfnamefont{G.}~\bibnamefont{De~Chiara}},
  \bibinfo{author}{\bibfnamefont{T.~J.~G.} \bibnamefont{Apollaro}},
  \bibinfo{author}{\bibfnamefont{F.}~\bibnamefont{Plastina}},
  \bibinfo{author}{\bibfnamefont{T.}~\bibnamefont{Busch}}, \bibnamefont{and}
  \bibinfo{author}{\bibfnamefont{M.}~\bibnamefont{Paternostro}},
  \bibinfo{journal}{New J. Phys.} \textbf{\bibinfo{volume}{15}},
  \bibinfo{pages}{043033} (\bibinfo{year}{2013}{\natexlab{b}}).

\bibitem[{\citenamefont{Francica et~al.}()\citenamefont{Francica, Apollaro, and
  et~al}}]{Francica}
\bibinfo{author}{\bibfnamefont{G.}~\bibnamefont{Francica}},
  \bibinfo{author}{\bibfnamefont{T.~J.~G.} \bibnamefont{Apollaro}},
  \bibnamefont{and} \bibinfo{author}{\bibnamefont{et~al}}, \bibinfo{note}{work
  in progress.}

\bibitem[{\citenamefont{Jozsa and Schumacher}(1994)}]{Josza94JMO41}
\bibinfo{author}{\bibfnamefont{R.}~\bibnamefont{Jozsa}} \bibnamefont{and}
  \bibinfo{author}{\bibfnamefont{B.}~\bibnamefont{Schumacher}},
  \bibinfo{journal}{J. Mod. Opt.} \textbf{\bibinfo{volume}{41}},
  \bibinfo{pages}{2343} (\bibinfo{year}{1994}).

\bibitem[{\citenamefont{Wootters}(1998)}]{Wootters98PRL80}
\bibinfo{author}{\bibfnamefont{W.~K.} \bibnamefont{Wootters}},
  \bibinfo{journal}{Phys. Rev. Lett.} \textbf{\bibinfo{volume}{80}},
  \bibinfo{pages}{2245} (\bibinfo{year}{1998}).

\bibitem[{\citenamefont{Amico et~al.}(2004)\citenamefont{Amico, Osterloh,
  Plastina, Fazio, and Massimo~Palma}}]{Amico04PRA69}
\bibinfo{author}{\bibfnamefont{L.}~\bibnamefont{Amico}},
  \bibinfo{author}{\bibfnamefont{A.}~\bibnamefont{Osterloh}},
  \bibinfo{author}{\bibfnamefont{F.}~\bibnamefont{Plastina}},
  \bibinfo{author}{\bibfnamefont{R.}~\bibnamefont{Fazio}}, \bibnamefont{and}
  \bibinfo{author}{\bibfnamefont{G.}~\bibnamefont{Massimo~Palma}},
  \bibinfo{journal}{Phys. Rev. A} \textbf{\bibinfo{volume}{69}},
  \bibinfo{pages}{022304} (\bibinfo{year}{2004}).

\bibitem[{\citenamefont{Apollaro et~al.}(2013)\citenamefont{Apollaro, Plastina,
  Banchi, Cuccoli, Vaia, Verrucchi, and Paternostro}}]{Apollaro13PRA88}
\bibinfo{author}{\bibfnamefont{T.~J.~G.} \bibnamefont{Apollaro}},
  \bibinfo{author}{\bibfnamefont{F.}~\bibnamefont{Plastina}},
  \bibinfo{author}{\bibfnamefont{L.}~\bibnamefont{Banchi}},
  \bibinfo{author}{\bibfnamefont{A.}~\bibnamefont{Cuccoli}},
  \bibinfo{author}{\bibfnamefont{R.}~\bibnamefont{Vaia}},
  \bibinfo{author}{\bibfnamefont{P.}~\bibnamefont{Verrucchi}},
  \bibnamefont{and}
  \bibinfo{author}{\bibfnamefont{M.}~\bibnamefont{Paternostro}},
  \bibinfo{journal}{Phys. Rev. A} \textbf{\bibinfo{volume}{88}},
  \bibinfo{pages}{052336} (\bibinfo{year}{2013}).

\bibitem[{\citenamefont{Apollaro et~al.}(2008)\citenamefont{Apollaro, Cuccoli,
  Fubini, Plastina, and Verrucchi}}]{Apollaro08PRA77}
\bibinfo{author}{\bibfnamefont{T.~J.} \bibnamefont{Apollaro}},
  \bibinfo{author}{\bibfnamefont{A.}~\bibnamefont{Cuccoli}},
  \bibinfo{author}{\bibfnamefont{A.}~\bibnamefont{Fubini}},
  \bibinfo{author}{\bibfnamefont{F.}~\bibnamefont{Plastina}}, \bibnamefont{and}
  \bibinfo{author}{\bibfnamefont{P.}~\bibnamefont{Verrucchi}},
  \bibinfo{journal}{Phys. Rev. A} \textbf{\bibinfo{volume}{77}},
  \bibinfo{pages}{062314} (\bibinfo{year}{2008}).

\end{thebibliography}
\end{document}